\documentclass[11pt]{article} 

\usepackage{latexsym}
\usepackage{amssymb}
\usepackage{amsmath}
\usepackage[breaklinks=true,linktocpage=true]{hyperref}
\usepackage{graphicx}
\renewcommand{\thefootnote}{\fnsymbol{footnote}}
\usepackage{mathbbol}

\newcommand{\tr}{{\rm Tr}}

\usepackage{xcolor}

\usepackage{cite}
\usepackage{ascmac}

\newcommand{\er}[1]{Eq.~\eqref{#1}}
\newcommand{\ers}[1]{Eqs.~\eqref{#1}}
\newcommand{\vev}[1]{\langle #1 \rangle}

\newcommand{\fr}{\frac}

\begin{document}
\pagestyle{empty}

\begin{flushright}
KEK-TH-2154
\end{flushright}

\vspace{3cm}

\begin{center}

{\bf\LARGE  
From 3d dualities to hadron physics
}
\\

\vspace*{1.5cm}
{\large 
Naoto Kan$^{1,2}$, Ryuichiro Kitano$^{1,2}$, 
Shimon Yankielowicz$^3$ and
Ryo Yokokura$^{1,4}$
} \\
\vspace*{0.5cm}

{\it 
$^1$KEK Theory Center, Tsukuba 305-0801,
Japan\\
$^2$Graduate University for Advanced Studies (Sokendai), Tsukuba
305-0801, Japan\\
$^3$The Raymond and Beverly Sackler School of Physics and Astronomy, 
Tel Aviv University, Ramat Aviv 69978, Israel
\\
$^4$Department of Physics \& Research and Education Center for Natural Sciences,
Keio University, Hiyoshi 4-1-1, Yokohama, Kanagawa 223-8521, Japan
}

\end{center}

\vspace*{1.0cm}

\begin{abstract}
{\normalsize
When one of the space-time dimension is compactified on $S^1$, the QCD
exhibits the chiral phase transition at some critical radius. When we
further turn on a background $\theta$ term which depends on the $S^1$
compactified coordinate, a topological ordered phase appears at low
energy via the winding of $\theta$. We discuss what kind of theories can
describe the physics near the critical point by requiring the matching
of topological field theories in the infrared.
As one of the possibilities, we propose a scenario where the $\rho$ and
$\omega$ mesons form a $U(N_f)$ gauge theory near the critical point.
In the phase where the chiral symmetry is restored, they become the dual
gauge boson of the gluon related by the level-rank duality between the
three dimensional gauge theories, $SU(N)_{N_f}$ and $U(N_f)_{-N}$.
}
\end{abstract} 

\newpage
\baselineskip=18pt
\setcounter{page}{2}
\pagestyle{plain}
\baselineskip=18pt
\pagestyle{plain}
 \setcounter{footnote}{0}%
\def\thefootnote{$*$\arabic{footnote}}%
   \def\@makefnmark{\hbox
       to\z@{$\m@th^{\@thefnmark}$\hss}}%

\section{Introduction}

The low energy limit of QCD is described by pions whose properties and
interactions have information of the global symmetry of QCD.
The lowest dimensional interaction terms can be determined once we know
the coset space of which the pions are the coordinate. Also, the
inconsistency in gauging a part of the global symmetry, {i.e., the
't~Hooft anomaly}, is encoded in the Wess-Zumino (WZ) term in the low
energy effective Lagrangian~\cite{Wess:1971yu,Witten:1983tw}.

The long distance behavior is also important near the critical point of
a phase transition. If the phase transition is smooth enough, one can
consider an effective theory of an order parameter which obtains a
vacuum expectation value (VEV) in the broken phase.
%
Global structures of QCD such as anomalies should also be kept in the effective theory for consistency of the theory.
Moreover, the 't~Hooft anomaly results in a matching conditions which
constrain the realization of the vacuum structure and the infrared
degrees of freedom~\cite{tHooft:1979rat,Frishman:1980dq,Coleman:1982yg}.
In the case of the finite temperature QCD above the QCD scale, such an
anomaly matching is usually trivially satisfied. The finite temperature
system can be regarded as an $S^1$ compactified QCD, and the effective
theory is, therefore, a three dimensional theory that has no chiral
anomaly.

The study of phase transitions in the three dimensional gauge theory has
a long
history~\cite{Banks:1977cc,Einhorn:1977jm,Peskin:1977kp,Horn:1979fy,Ukawa:1979yv,Dasgupta:1981zz,Appelquist:1981vg,Gopfert:1981er,Appelquist:1986fd,Appelquist:1988sr,Appelquist:1989tc}
with some important recent
developments~\cite{Closset:2012vp,Aharony:2015mjs,Karch:2016aux,Karch:2016sxi,Seiberg:2016gmd,Gomis:2017ixy,Freed:2017rlk,Komargodski:2017keh,Gaiotto:2017tne,Benini:2017dus,Choi:2018tuh,Komargodski:2018odf,Bashmakov:2018wts,Armoni:2019lgb,Argurio:2018uup}.
Even though there is no chiral anomaly, there are topological orders in
the low energy effective theories. For example, $SU(N)$ gauge theory can
have the CS term with an integer level, $k$. The CS term dominates the
infrared physics, and it reduces to the CS theory, denoted as $SU(N)_k$,
which has a gap, but the Wilson lines have non-trivial values depending
on their topology.
This non-trivial topological behavior should be matched when we discuss
the effective description near the phase transition.
Based on this discussion, the dualities between CS theories coupled to
fermions and those coupled to bosons have been proposed and checked. The
precise forms of the dualities are listed in
Ref.~\cite{Aharony:2015mjs}.
Based on these dualities, the phase diagrams of the three dimensional
QCD (QCD$_3$) have been discussed~\cite{Komargodski:2017keh}.
In particular, it has been conjectured that the $SU(N)_{k}$ theory with
$N_f$ $(> 2|k|)$ fermions undergoes the symmetry breaking $U(N_f) \to
U(N_f/2 + k) \times U(N_f/2 - k)$ when the fermions masses are smaller
than some critical value.
There is a phase transition between the symmetry broken and unbroken
phases as the fermion masses are varied.
If the transition is of the second order, near the critical point, there
is a dual description of the theory by the $U(N_f/2 + k)_{-N}$ or
$U(N_f/2 - k)_{N}$ theory coupled to $N_f$ scalar fields whose Higgs
phenomenon describes the phase transition.

In the unbroken phase in QCD$_3$ the CS theory describes the low energy
physics. The fermionic theory flows to the $SU(N)_{\pm N_f/2 + k}$
theory while it is the $U(N_f/2 \pm k)_{-N}$ theory in the bosonic
theory. These two theories are related by the level-rank
duality~\cite{Moore:1989yh,Naculich:1990pa,Mlawer:1990uv,Naculich:2007nc}
and are known to give the same physics~\cite{Hsin:2016blu}.  It is quite
not trivial that the matching of the low energy limit is realized in
this way.

In this paper, motivated by the symmetry breaking and its dual
description in QCD$_3$, we discuss the low energy limits of the $S^1$
compactified QCD in four space-time
dimensions~\cite{Unsal:2007vu,Unsal:2007jx,Unsal:2008ch,Shifman:2008cx,Shifman:2008ja,Shifman:2009tp,Cossu:2009sq,Myers:2009df,Unsal:2010qh,Simic:2010sv,Thomas:2011ee,Anber:2011gn,Poppitz:2012nz,Unsal:2012zj,Argyres:2012ka,Anber:2013doa,Cossu:2013ora,Bhoonah:2014gpa,Anber:2014lba,Bergner:2014dua,Misumi:2014raa,Li:2014lza,Iritani:2015ara,Anber:2015kea,Anber:2015wha,Cherman:2016hcd,Anber:2017rch,Komargodski:2017keh,Cherman:2017dwt,Tanizaki:2017mtm,Tanizaki:2017qhf,Aitken:2017ayq,Ivanova:2018grj,Hongo:2018rpy,Anber:2018xek}
with the hope that the three dimensional duality may give us some hints
of the four dimensional physics.
Indeed for abelian gauge theories it has been pointed out that the three
dimensional duality is lifted up to the S-duality in the four
dimensional theory~\cite{Aitken:2018joz}.
We consider a background $\theta$ term which depends on the coordinate
of the $S^1$ direction. In particular, the function $\theta$ can have
windings along the $S^1$
direction~\cite{Cordova:2019jnf,Cordova:2019uob}, which determines the
CS level in the three dimensional effective theory.
As the most interesting example, one can take the background with the
winding number, $N_f$, which is the number of quark flavors in four
dimensions.
For a small radius, one finds that the low energy theory is
$SU(N)_{N_f}$ where the vacuum is gapped. At a large radius, the theory
is better described by the chiral Lagrangian for pions with the WZ
terms. Since the low energy limits are different, there must be a phase
transition at some critical radius.
The background $\theta$ induces a winding of the $\eta'$ meson, 
which gives rise to the non-trivial WZ
term in the three-dimensional theory.
The two limits of the theory can be interpolated by the Higgs mechanism
of the $U(N_f)_{-N}$ theory coupled to $2N_f$ flavors of scalar
fields. Here, again, in the unbroken phase, the theory is related by
the level-rank duality.
If this picture is correct, the natural candidate for this dual $U(N_f)$
gauge bosons are the $\rho$ and the $\omega$ mesons (in the spirit of
Ref.~\cite{Sakurai:1960ju}), which we know, phenomenologically, to be
successfully described in terms of gauge
fields~\cite{Bando:1984ej,Bando:1985rf,Bando:1987br,Son:2003et,Sakai:2004cn,Erlich:2005qh,DaRold:2005mxj}. See
also~\cite{Komargodski:2010mc, Kitano:2011zk, Abel:2012un} for the
interpretation of the Seiberg duality as the gauge theory of vector
mesons, and Refs.~\cite{Armoni:2017jkl, Akhond:2019ued} for the relation
between the Seiberg duality and the phase structure of QCD$_3$.
The ``gauge bosons'' are massive anyway by the CS term even in the
unbroken phase in the three dimensional effective theory. Therefore,
they are not quite effective degrees of freedom, but their existence is
important for having a non-trivial topological order required from the
matching of the low energy physics.

We discuss the similarities between the QCD$_3$ for small $k$ and
QCD$_4$, and propose an exotic possibility that the $\rho$ and $\omega$
mesons continuously becoming a dual gauge boson of gluons related by the
level-rank duality. This conjectured picture is at least consistent when
the winding of $\theta$ is less than $N_f$, especially, when it is zero
at which case the system can be regarded as the finite temperature QCD.

The paper is organized as follows: In section \ref{sec:qcd3} we discuss
the pattern of the flavor symmetry breaking in QCD$_3$ with $2N_f$
fermions and its breaking to the chiral symmetry group associated with
QCD$_4$. We identify the order parameter which leaves the correct
Nambu-Goldstone bosons massless to match the one of QCD$_4$. The
corresponding effective low energy theory is written down with an
emphasis on the CS terms and the related WZ terms, also in the presence
of external gauge fields. In particular, we identify the term associated
with baryon number.
These terms are related to the flavor anomalies in QCD$_4$. They were
recently discussed also by Komargodski in~\cite{Komargodski:2018odf}. In
section~\ref{sec:qcd4} we put QCD$_4$ on $M^3 \times S^1$ including a
$\theta$ angle which winds along $S^1$ and background gauge fields which
depend on the $S^1$ coordinate. We explore the theory for small radius
($\Lambda_4 R \ll 1$) and large radius ($\Lambda_4 R \gg 1$) and argue
for the existence of a phase transition at some critical $R_*$.
In section~\ref{sec:hadron} we speculate/conjecture on the possible
behavior of the hadronic vector mesons ($\rho$, $\omega$, ...) near the
critical point and the nature of the theory at the critical point. In
particular, we put forward a scenario in which the hadronic vector
mesons become gauge bosons and give rise to an $U(N_f)$ gauge theory at
the critical point.
In section~\ref{sec:holographic} we present holographic QCD-like models
represented by quiver diagrams which capture this conjectured scenario
of vector mesons as gauge bosons. In section~\ref{sec:finiteT} we
comment briefly on the implications of our study on finite temperature
QCD and possible scenarios for the nature of the phase transition at
$T_*=1/R_*$. Lattice simulations may decide between these various
scenarios. 
Section~\ref{sec:discussion} is devoted to discussion.
There are three appendices. In Appendix~\ref{sec:quiver} we write down
the Lagrangian associated with the quiver diagram in
section~\ref{sec:holographic}.
Appendix~\ref{sec:windingTheta} addresses the issue of the integration
over $S^1$ when a winding $\theta$ term is present.
In Appendix~\ref{sec:baryon} we discuss the issue of the Baryon number
and the (winding) configuration of $\eta'$ by
considering the WZ term and the associated anomaly of chiral $U(1)_A$
current, $U(1)_{\rm EM}$ electromagnetic current and the $U(1)$ baryon
number. In particular we can consider the situation that $\eta' $ winding
happens on a finite sheet and localized on $S^1$. The sheet
configuration is just the Hall droplet studied in
\cite{Komargodski:2018odf}. Following \cite{Komargodski:2018odf} we can
identify the Baryonic configuration which resides on the boundary of the
finite region. It is ``amusing'' to note that one can also identify a
non-local configuration which corresponds to the quark with its Wilson
line going into the bulk. In this sense the quark appears as a
``soliton'' in the hadronic effective theory.

\section{QCD$_3$ and chiral symmetry breaking}
\label{sec:qcd3}

Based on the studies of the three dimensional dualities, it has been
conjectured that the low energy theory of three dimensional $SU(N)_0$
QCD with $2N_f$ fermions is described by a non-linear sigma model with
the target space,
\begin{align}
 {U(2N_f) \over
U(N_f) \times U(N_f)}
\label{eq:nlsm}
\end{align}
for small enough fermion
masses~\cite{Vafa:1984xh,Komargodski:2017keh}. There is an upper bound
on $N_f$ although the precise location is
unknown~\cite{Sharon:2018apk}. It is noted that an appropriate WZ term
should be added in the Lagrangian. The same low energy theory is
obtained by a linear sigma model with $U (N_f)_N$ or $U(N_f)_{-N}$ gauge
theory coupled to $2 N_f$ scalar fields.
Beyond the critical value of the fermion mass, both the fermionic and
bosonic theories flow to topological field theories related by the
level-rank duality, $SU(N)_{\pm N_f} \leftrightarrow U(N_f)_{\mp N}$.

We would like to discuss the relation between this symmetry breaking
phenomena and the chiral symmetry breaking in QCD$_4$.
In QCD$_4$ with $N_f$ Dirac fermions (corresponding to $2N_f$ fermions
in QCD$_3$), the low energy theory is a non-linear sigma model with
\begin{align}
 {SU(N_f)_L \times SU(N_f)_R \over SU(N_f)_{L+R}}.&
\label{eq:coset}
\end{align}
This coset space is a subspace of \eqref{eq:nlsm}. 
One can reduce the
space \eqref{eq:nlsm} to \eqref{eq:coset} by adding an explicit breaking
terms in the Lagrangian.

In this section, we investigate and discuss the physics of QCD$_3$ 
deformed by an explicit breaking term
to exhibit a similar symmetry breaking pattern as QCD$_4$.
Throughout this section, we stay in three-dimensional spacetime, 
and analyze the phase structure near the critical quark mass.
Later, in section \ref{sec:qcd4} and \ref{sec:hadron}, 
we will compare and relate 
the phase structure of the deformed QCD$_3$ revealed in this section,
to the critical phase transition in QCD$_4$ compactified on a circle.
\footnote{
Later in this section, we will comment on the relation of a 
deformed version of QCD$_3$ to QCD$_4$ compactified on a circle.
This will allow us to borrow and use some techniques which are familiar in QCD$_4$ for the investigation of the deformed QCD$_3$.
}

In what follows, we explicitly break the $U(2N_f)$ symmetry 
of QCD$_3$ by hand,
$U(2N_f) \to U(N_f)_L \times U(N_f)_R$, and then discuss the spontaneous
breaking $U(N_f)_L \times U(N_f)_R \to U(N_f)_{L+R}$.  
We denote the $N_f + N_f$ flavors of the QCD$_3$ 
by $\psi$ and $\tilde \psi$, and
 introduce the explicit breaking term by coupling a massive
adjoint scalar field $a_3$ to QCD$_3$,
\begin{align}
 {\cal L}_{a_3} = - \bar \psi a_3 \psi + \bar {\tilde \psi} a_3 {\tilde \psi},
\label{eq:explmass}
\end{align}
this interaction breaks the $U(2N_f)$ symmetry to $U(N_f) \times
U(N_f)$. 
In QCD$_4$, the role of $a_3$ is played by the extra component
of the gauge boson.

If we do not introduce the explicit breaking term in \er{eq:explmass},
the symmetry breaking pattern is described in Eq.~\eqref{eq:nlsm}.
This breaking pattern suggests that the
order parameter is
\begin{align}
 \langle \bar \psi \psi - \bar {\tilde \psi} \tilde \psi \rangle
\label{eq:vectorVEV}
\end{align}
and/or
\begin{align}
 \langle \bar \psi \tilde \psi \rangle.&
\label{eq:bartilde}
\end{align}
Both VEVs reproduce the symmetry pattern $U(2N_f) \to U(N_f) \times
U(N_f)$ as they are actually equivalent by $U(2N_f)$ flavor rotation.

Once the $U(2N_f)$ breaking terms in Eq.~\eqref{eq:explmass} are
introduced, as long as $a_3$ is heavy enough, at least one of the VEVs
in \er{eq:vectorVEV} and \er{eq:bartilde} should remain non-vanishing.
They are not anymore equivalent, and it is a dynamical issue which VEV
remains non-vanishing.
Indeed, the symmetry breaking pattern is different. If $\langle \bar
\psi \tilde \psi \rangle$ is non-vanishing, the $U(N_f) \times U(N_f)$
symmetry is spontaneously broken down to $U(N_f)$ while there is no
symmetry breaking if it vanishes, as $ \langle \bar \psi \psi - \bar
{\tilde \psi} \tilde \psi \rangle$ is invariant under $U(N_f) \times
U(N_f)$.

We argue that $ \langle \bar \psi \psi - \bar {\tilde \psi} \tilde \psi
\rangle = 0$ should be chosen at the massless point and thus for small
fermion masses the symmetry breaking pattern is $U(N_f) \times U(N_f)
\to U(N_f)$.
One can see the symmetry 
breaking pattern by analyzing the effective Lagrangian 
after integrating out the massive scalar field $a_3$.
In the classical level, 
the effective Lagrangian can be evaluated as
\begin{equation}
 {\cal L}_{{\rm eff.},a_3} 
 = -\fr{1}{8M^2}
((\bar{\psi}\psi)^2 + (\bar{\tilde{\psi}}\tilde{\psi})^2 
- 2 |\bar\psi \tilde{\psi}|^2 
)
-\fr{1}{4 M^2 N}(\bar\psi \psi - \bar{\tilde{\psi}}\tilde{\psi})^2
+ \cdots,
\label{a3intout}
\end{equation}
where $M$ is the mass of $a_3$, 
and the ellipsis ``$\cdots$'' denotes terms given by
$(\bar\psi \gamma^\mu \psi )(\bar\psi \gamma_\mu \psi)$,
$(\bar{\tilde{\psi}} \gamma^\mu \tilde\psi ) (\bar{\tilde{\psi}}
\gamma_\mu \tilde\psi)$, and $(\bar{\psi} \gamma^\mu \tilde\psi )
(\bar{\tilde{\psi}} \gamma_\mu\psi)$ .  The effective potential in
\er{a3intout} implies that the potential energy increases if the VEV in
\er{eq:vectorVEV} is non-zero, and decreases if the VEV in
\er{eq:bartilde} is non-zero under the assumption that
$\vev{(\bar{\psi}\psi)^2 }$, $\vev{(\bar{\tilde{\psi}}\tilde{\psi})^2
}$, and $\vev{(\bar{\psi}\tilde{\psi})^2 }$ can be approximately written
as $\vev{\bar{\psi}\psi}^2 $, $\vev{\bar{\tilde{\psi}}\tilde{\psi} }^2$,
and $\vev{(\bar{\psi}\tilde{\psi}) }^2$, respectively.

The discussion in the last paragraph lacks rigor.
To make a more precise argument, we use the fact that QCD$_3$ 
deformed by \er{eq:explmass} can be obtained as the lowest Kaluza-Klein (KK) mode of QCD$_4$ compactified on a circle
in a meta-stable vacuum in a weakly-coupled regime, 
namely at a small radius. 
In this weakly-coupled regime, we can control the validity of our approach.
More precisely, the three dimensional model with the explicit
breaking term in \er{eq:explmass} can be obtained as the low energy
effective theory of the $S^1$ compactified QCD$_4$ with $N_f$ Dirac
fermions, $\Psi$, as we will see in the next section. 
For the current discussion, one can take the $\theta$ term in
QCD$_4$ to be absent in order to match to the Chern-Simons (CS) level, 
$k=0$,
in QCD$_3$. We will discuss the case with non-zero $\theta$ term in the next
section.
As is shown in Fig.~\ref{fig:potential} of section \ref{sec:qcd4}, 
the model with massless fermions can be realized at a meta-stable vacuum
where the Wilson loop is trivial, $\langle e^{i \int a_3} \rangle = {\bf
1}$, when $N_f < N$. The periodic boundary condition is taken for
$\Psi$.\footnote{Note that the fermions become massive in the true vacuum.}
In this language, $ \langle \bar \psi \psi - \bar {\tilde \psi} \tilde
\psi \rangle$ and $\langle \bar \psi \tilde \psi \rangle$ correspond to
$\langle \bar \Psi \gamma^3 \Psi \rangle$ and $\langle \bar \Psi \Psi
\rangle$, respectively, where $x_3$ is the $S^1$ direction.

Let us consider the Euclidean theory on a torus (which has no 
$\theta$ term) for which the path integral measure is non-negative.
We will further take the periodic boundary condition for $\Psi$ along the $x_4$
direction in addition to the $x_3$ direction.
If
there is a meta-stable 
vacuum with $\langle \bar \Psi \gamma_3 \Psi \rangle \neq 0$,
there should also be 
a meta-stable vacuum with 
a
non-vanishing fermion
density,
$\langle \bar \Psi \gamma_4 \Psi \rangle \neq 0$.
This follows from the fact
 that this Euclidean theory with radii $(R_3, R_4)$ along 
$(x_3, x_4)$ directions is equivalent to the one with radii 
$(R_4, R_3)$ along $(x_3, x_4)$, since we take the same periodic boundary 
conditions along these directions.

Let us assume that $\langle \bar \Psi \gamma_4 \Psi \rangle \neq 0$
in the vacuum, and show that it leads to a contradiction.
The following discussion is essentially identical to the Vafa-Witten
theorem~\cite{Vafa:1983tf,Vafa:1984xg}.
The non-zero $\vev{\bar{\Psi} \gamma_4 \Psi}$ 
corresponds to non-zero charge density 
$\vev{\bar{\Psi} \gamma_0 \Psi}$
in the Minkowskian spacetime.
We will deform the Minkowskian action by adding 
the term $\mu \bar{\Psi} \gamma_0 \Psi$.
Here, $\mu$ is a real chemical potential,
and we assume that $|\mu|$ is sufficiently small so that 
higher-order perturbations of the free energy in $\mu$ 
can be neglected, and the meta-stability of the vacua is not violated.
Note that the VEV $\vev{\bar{\Psi} \gamma_0 \Psi}$
is non-zero
in the limit $\mu \to 0$ by assumption.
In the presence of a small but non-zero $\mu$, 
the free energy becomes 
$\mu \vev{\bar{\Psi} \gamma_0 \Psi} + E_0$
to lowest order in $\mu$,
where $E_0$ is the free energy at $\mu=0$.
In general, $E_0$ depends on the VEV of the Wilson loop
which labels meta-stable and stable vacua.
Since $\vev{\bar{\Psi} \gamma_0 \Psi}$ 
can be positive or negative, there is a vacuum where 
$\mu \vev{\bar{\Psi} \gamma_0 \Psi}$ is negative,
hence lowering the energy.

Now, we show that the presence of the non-zero VEV 
$ \vev{\bar{\Psi} \gamma_0 \Psi}$ is impossible 
by using the expression in the Euclidean path integral.
The term $\bar{\Psi} \gamma_0 \Psi$
in the Minkowskian spacetime 
corresponds to $\bar{\Psi} \gamma_4 \Psi$
in the Euclidean space.
The above deformation by the chemical potential 
in the Minkowski spacetime
corresponds to the deformation 
by adding $\mu \bar{\Psi} \gamma_4 \Psi$
in the Euclidean action.
In the Euclidean QCD$_4$, however,
since the path integral measure is
non-negative~\cite{Vafa:1983tf} 
while adding a chemical potential gives a phase, the free
energy 
cannot decrease with $\mu$
and has (for small $\mu$) its minimal value
at vanishing chemical potential, i.e., $\mu =0$.
Therefore, having a non-zero VEV 
$ \vev{\bar{\Psi} \gamma_4 \Psi}$
contradicts the non-negativity of the 
path integral.
This means $ \vev{\bar{\Psi} \gamma_4 \Psi}=0$
for any radius of the $x_4$ direction,
implying that also $\vev{\bar{\Psi} \gamma_3 \Psi} =0$,
in contradiction to our assumption $\vev{\bar{\Psi} \gamma_3 \Psi}\neq 0$.

%

We emphasize that the conclusion holds when we restrict the path
integral to the background with $\langle e^{i \int a_3} \rangle = {\bf
1}$ which leads to a meta-stable vacuum.
Our argument relies only on the non-negative measure of the path 
integral which holds in any Euclidean background of the gauge field.
This enables us to analyze the the vacuum structure of QCD$_3$ and
conclude that QCD$_3$ with massless $2N_f$ flavors with no CS term ($k=0$)
(thus having a non-negative measure) and explicit breaking term is
realized as the low energy effective theory of QCD$_4$ with $N_f$ flavors
with $\theta=0$ compactified on $S^1$ at a meta-stable vacuum.

Next, we consider a limit to match to the three dimensional theory. 
By the KK expansion of $\langle \bar \Psi \gamma^3 \Psi
\rangle $, the integral of the VEV along the $S^1$ direction is given by
\begin{equation}
\int^{2\pi R}_{0} d x_3 \langle \bar \Psi \gamma^3 \Psi \rangle  = 
 \vev{\bar\psi \psi - \bar{\tilde\psi}\tilde\psi}
 + \sum_{n \neq 0} \vev{\bar\psi_{n} \psi_n - \bar{\tilde\psi}_{n}\tilde\psi_n},
\end{equation}
where $\psi_n$ and $\tilde\psi_n$ denote the $n$-th KK modes.  The VEV
$\vev{\bar\psi \psi - \bar{\tilde\psi}\tilde\psi}$ in the three
dimensional theory corresponds to one of the lowest $(n=0)$ KK mode.
As we will see in section \ref{sec:qcd4}, one can have a finite energy
gap between the mass scale of the first KK mode, $1/R$, and the
dynamical scale of QCD$_3$ when the theory is weakly coupled at the
compactification scale. Therefore, there exists a limit where higher KK
modes are decoupled, and only the zero modes participate in the three
dimensional dynamics.  In this limit, we have 
$\int^{2\pi R}_{0} d x_3 \langle \bar \Psi \gamma^3 \Psi \rangle = \vev{\bar\psi \psi - \bar{\tilde\psi}\tilde\psi}$.
By using 
$ \langle \bar \Psi \gamma^3 \Psi \rangle =0$, we obtain
\begin{equation}
 \vev{\bar\psi \psi - \bar{\tilde\psi}\tilde\psi} =0.
\end{equation}
Therefore, in the three-dimension limit, $\langle \bar \psi \tilde \psi
\rangle \neq 0$ should be chosen at the massless point and the same
symmetry breaking pattern is expected for small fermion masses.

For $\langle \bar \psi \tilde \psi \rangle \neq 0$
in the presence of the explicit breaking terms
in \er{eq:explmass}
, the $U(N_f) \times
U(N_f)$ chiral symmetry is broken to $U(N_f)$, leaving a part of the
Nambu-Goldstone modes massless and matches to the QCD$_4$ up to an
anomalous axial $U(1)_A$
\footnote{
In QCD$_4$, the axial $U(1)_A$ symmetry is also broken 
by the axial anomaly. The corresponding Nambu-Goldstone mode, 
the $\eta'$ meson, becomes massive and is decoupled from the 
low-energy effective theory. 
In three dimensions, we can explicitly break 
the $U(1)_A$ symmetry by adding 
$|\log \det \bar{\tilde \psi} \psi |^2$, 
which corresponds to $ | \log \det \xi|^2 $
in the low-energy effective theory.
Since the $\eta'$ meson becomes massless in the large $N$ limit,
our discussion is valid at large $N$.
}.
%
Therefore, it is possible that the chiral symmetry breaking in QCD$_4$
has something to do with the phase transition in QCD$_3$ where there is
a dual description by the Higgs mechanism of $U(N_f)_{\pm N}$ gauge
theory.
In QCD$_4$, it is well-known that the physics of the vector mesons,
$\rho$ and $\omega$, is nicely described by the color-flavor locked
phase of $U(N_f)$ gauge theory.
The structure of the dual bosonic theory
in three dimension is indeed of this type as is revealed in the 
following discussion based on the low energy effective theory.

Let us look at the low energy effective theory. We first discuss the
theory without the explicit breaking terms in Eq.~\eqref{eq:explmass}.
In this case, the non-zero VEV $\vev{\bar\psi \tilde\psi}$ breaks 
$U(2N_f) $ to $U(N_f) \times U(N_f)$.
In the dual $U(N_f)_{\pm N}$ bosonic theory, this symmetry breaking
corresponds to a color-flavor locked phase:
the $U(N_f)$ gauge symmetry is completely Higgsed, and the $U(2N_f)$
flavor symmetry is broken, while a part of $U(2N_f)$ flavor rotation
together with a $U(N_f)$ gauge rotation is unbroken.
The unbroken flavor symmetry is $U(N_f) \times U(N_f)$.
%
The pions in this symmetry breaking have the WZ term as discussed in
Ref.~\cite{Komargodski:2017keh}. The WZ term can be obtained from the
CS term in the bosonic $U(N_f)_{\pm N}$ theory.
The uneaten $2N_f^2$ Nambu-Goldstone fields in Eq.~\eqref{eq:nlsm} are
introduced as a $2N_f \times 2N_f$ matrix:
\begin{align}
 \xi& = \exp \left[
{{i \over \sqrt 2}  \pi^a \left(
\begin{array}{cc}
 0 & T^a \\
 T^a & 0 \\
\end{array}
 \right)}
+ {{i \over \sqrt 2} \tilde \pi^a \left(
\begin{array}{cc}
 0 & i T^a \\
-i T^a & 0 \\
\end{array}
 \right)}
 \right],
\end{align}
where $T^a$ are the generators of $U(N_f)$ group.  The Higgsed $U(N_f)$
gauge field, $b_\mu$ couples to $\xi$ as
\begin{align}
 {\cal L}_{\rm NLSM}& = {f^2 \over 4} \left|
 \partial_\mu \xi + i \xi \left(
\begin{array}{cc}
 {b}_\mu &  0\\
 0 & 0 \\
\end{array}
 \right)
 \right|^2.
\end{align}
In the low-energy limit, the gauge field $b_\mu$ can be integrated out,
since this gauge field is massive by the Higgs mechanism.
The equation of motion for $b_\mu$ gives
\begin{align}
 {b}_{ij}& = \left( \xi^{-1} d \xi \right)_{ij},
\end{align}
where $i,j = 1, \cdots, N_f$ run the first half of the $2N_f$
indices. Substituting this into the CS term,
\begin{align}
 {S}_{\rm CS}& =  \pm {N \over 4 \pi} \int_{M^3} {\rm Tr} \left(
 {b} d {b} + {2 \over 3} {b}^3
 \right),
\end{align}
one obtains the WZ terms.

We now introduce the explicit breaking term in \er{eq:explmass}.
A half of pions obtain masses when we include the interaction to break
the $U(2N_f)$ symmetry to the chiral symmetry. For example, one can
introduce a spurion field,
\begin{align}
 X = \left(
\begin{array}{cc}
 0 & {\bf 1} \\
 {\bf 1} & 0 \\
\end{array}
\right),
\end{align}
and write down a $U(2N_f)$ breaking term,
\begin{align}
 {\rm Tr} \left(
\xi^{-1} X \xi X
\right).
\end{align}
This term gives a mass to $\tilde \pi$ while leaving $\pi$ massless.

In the low-energy limit where $\tilde\pi^a$ are decoupled, one can set
$\tilde \pi^a = 0$, and the WZ term among pions vanishes. However, one
can trace the existence of the WZ term by turning on external gauge
fields.
Let us introduce, the background gauge fields for the unbroken $U(N_f)$
global symmetry $A^\mu$, as the one which couples to the
$U(N_f)$ vector current, $\bar \psi \gamma^\mu T^a \psi + \bar {\tilde
\psi} \gamma^\mu T^a {\tilde \psi}$. The $U(1)$ part is the baryon
number. The equation of motion for ${b}$ now gives, ${b} =
A + \cdots$, and thus we have
\begin{align}
 {S}_{\rm WZ}& =  \pm {N \over 4 \pi} \int_{M^3} {\rm Tr} \left(
 A dA + {2 \over 3} A^3
 \right) + \cdots .
\label{eq:WZext}
\end{align}
In particular, we have a term
\begin{align}
 \pm {1 \over 4 \pi} \int_{M^3} B {\rm Tr} (dA),
\end{align}
where $B$ is the baryon number normalized such that the quarks have the
charge $1/N$.
The above term gives the baryon number $(B=1)$ for a monopole that has
the unit magnetic charge of the $U(1)$ subgroup of $U(N_f)$. For
example, the monopole made of the (11)~component of $b$ has the baryon
number $B=1$.

\section{QCD$_4$ on a circle}
\label{sec:qcd4}

We discuss the four-dimensional QCD compactified on $S^1$ and look for a
relation to the phase transition in QCD$_3$.
We start with the action of $SU(N)$ gauge theory coupled with massless
$N_f$ Dirac fermions, $\Psi_i$, $(i=1,\cdots,N_f)$, 
with the assumption $N > N_f$
on $M^3 \times S^1$,
\begin{align}
 S = \int_{M^3 \times S^1} d^4 x
\biggl[
&
-{1 \over 2 g_4^2} {\rm Tr} \left(f_{M N} f^{M N} \right)
+ {\theta (x_3) \over 32 \pi^2} \epsilon_{MNPQ} 
{\rm Tr} \left( f^{MN} f^{PQ} \right)
\nonumber \\
&
+ i \bar \Psi_i \gamma^M (\partial_M - i a_M) \Psi_i
\nonumber \\
&
- \partial_M \alpha_L^i (x_3) \bar \Psi_i \gamma^M P_L \Psi_i
- \partial_M \alpha_R^i (x_3) \bar \Psi_i \gamma^M P_R \Psi_i
\bigg].
\label{eq:qcd}
\end{align}
The periodic boundary conditions are imposed on gauge fields. The
Lorentz indices, $M$, $N$, $P$, $Q$, run from $0$ to $3$, where $x_3$ is
the $S^1$ direction. $P_{L,R}$ are projection operators of chirality,
$P_{L,R} = (1 \mp \gamma_5)/2$.
We introduced $x_3$ dependent background fields, $\theta (x_3)$ and
$\alpha^i_{L,R} (x_3)$.
The boundary condition of the quarks are
\begin{align}
 \Psi_i (x_3 + 2 \pi R) = e^{i \nu} \Psi_i (x_3),
\end{align}
where $0 \leq \nu < 2 \pi$.

The $S^1$ compactification requires that $\theta$ and $\alpha$ are also
valued on $S^1$ that allows 
\begin{align}
 \int_{S^1} d\theta = 2 \pi k, \quad
 \int_{S^1} d \alpha_{L,R}^i = 2 \pi m^i_{L,R},
 \label{eq:periodicities}
\end{align}
where $k$ and $m^i_{L,R}$ are integers.
The integral on $S^1$ should be properly defined as in
Ref.~\cite{Cordova:2019jnf} so that the partition function does not
depend on the coordinate system on $S^1$. (See
Appendix~\ref{sec:windingTheta} for the definition.)
There is a redundancy due to the anomalous chiral symmetry, $\Psi_{L(R)i} \to
e^{i \beta^i_{L(R)}} \Psi_{L(R)i}$,
\begin{align}
 \theta (x_3) & \to \theta (x_3) - \sum_i (\beta^i_R (x_3) - \beta^i_L (x_3)), \quad
 \alpha^i_{L,R} (x_3) \to \alpha^i_{L,R} (x_3) + \beta^i_{L,R} (x_3),
\label{eq:redundancy}
\end{align}
where 
\begin{align}
 \int_{S^1} d\beta_{L,R}^i  = 2 p \pi,
\end{align}
with $p$ integers to maintain the boundary conditions.

In the following discussion, we are particularly interested in the theory with
\begin{align}
k = N_f, \quad m_{L,R}^i = 0.
\label{eq:param}
\end{align}
since the vacuum structure looks the same as the three dimensional case
discussed in the previous section.
By the chiral rotations, this theory is equivalent to, for example,
\begin{align}
k = 0, \quad m_R^i = 1, \quad m_L^i = 0.
\label{eq:thetazero}
\end{align}
The physics should depend on the combination:
\begin{align}
\bar \theta (x_3) =  \theta (x_3) + \sum_i (\alpha^i_R (x_3) - \alpha^i_L (x_3)),
 \quad \bar k& = k + \sum_i (m_R^i - m_L^i).
 \label{eq:combi}
\end{align}

We discuss the phase structure of the theory as a function of the radius $R$.
The dynamical scale $\Lambda_4$ is defined as
\begin{align}
 \Lambda_4^b& = \Lambda^b e^{-{8 \pi^2}/g_4^2 (\Lambda)},\quad
b = {11 \over 3} N - {2 \over 3} N_f,
\end{align}
where $\Lambda$ is an arbitrarily high scale. 
For $\Lambda_4 R \gg 1$, the low energy dynamics is described by hadrons
on an $S^1$ compactified background.
In the other limit, $\Lambda_4 R \ll 1$, the low energy description is a
three dimensional gauge theory on $M^3$ via the KK
decomposition.  The gauge coupling constant in the three dimensional
effective theory is given by
\begin{align}
 {1 \over g_3^2}& = {2 \pi R \over g_4^2 (1/R)} ,
\end{align}
and the dynamical scale in the three dimensional theory is defined as
\begin{align}
 \Lambda_3& = {g_3^2 N \over 8 \pi} = {g_4^2 (1/R) N \over 16 \pi^2 R}
= {1 \over R} \left(
- {2 b \over N} \log (\Lambda_4 R)
\right)^{-1}.
\end{align}
One can see that for a small enough $\Lambda_4 R$, there is an energy
gap between the dynamical scale $\Lambda_3$ and the mass of the first KK
mode, $1/R$.

\subsection{Small radius, $\Lambda_4 R \ll 1$}
Let us use the basis with $k=0$.
The KK expansion of the fermions can be done as
\begin{align}
 \Psi (x, x_3)& = \sum_{n = - \infty}^\infty \left(
\begin{array}{c}
 \psi_n (x) \phi_n (x_3) \\
 \sigma^3 \tilde \psi_n (x) \tilde \phi_n (x_3) \\
\end{array}
\right),
\end{align}
with
\begin{align}
\phi_n (x_3) = {1 \over \sqrt {2 \pi R}} 
\exp \left[
i \alpha_L (x_3)
+ i \left( {n \over R} + {m_L \over R} + {\nu \over 2 \pi R} \right) x_3
\right],
\end{align}
and
\begin{align}
\tilde \phi_n (x_3) = {1 \over \sqrt {2 \pi R}} 
\exp \left[
i \alpha_R (x_3)
+ i \left( {n \over R} + {m_R \over R} + {\nu \over 2 \pi R} \right) x_3
\right].
\end{align}
The three dimensional effective action is given by 
\begin{align}
 S_{\rm eff} = \int d^3 x
\bigg [ 
&
- {1 \over 2 g_3^2} {\rm Tr} \left(
f_{\mu \nu} f^{\mu \nu}
\right)
\nonumber \\
& 
+ \sum_n i \bar \psi_n \gamma^\mu (\partial_\mu - i a_\mu) \psi_n 
+ \sum_n i \bar {\tilde \psi}_n \gamma^\mu (\partial_\mu - i a_\mu)
 \tilde \psi_n 
\nonumber \\
&
- \sum_n 
 \left( -{m_L \over R} - {n \over R} - {\nu \over 2 \pi R} \right) \bar \psi_n \psi_n
\nonumber \\
&
- \sum_n 
 \left( {m_R \over R} + {n \over R} + {\nu \over 2 \pi R}   \right)
 \bar {\tilde \psi}_n \tilde \psi_n 
\nonumber \\
&
+ {1 \over  g_3^2} {\rm Tr} (D_\mu a_3 D^\mu a_3)
- (\bar \psi a_3 \psi - \bar {\tilde \psi} a_3 \tilde \psi )
- V(a_3)
\bigg ].
\label{eq:3d-eff}
\end{align}
Here, $V(a_3)$ is the effective potential for the gauge field along 
the $S^1$ direction, which we will determine in \er{eq:general_one_loop}.
We have dropped the massive KK modes of gauge fields. 
At this stage, one can see that the effects of $m_L$ and $m_R$ are to
shift the KK spectrum of the fermions by $1/R$, and thus can be
absorbed by the redefinition of $n$. However, in three dimensions, the
signs of the fermion masses are important, and thus one cannot simply
ignore $m_{L,R}$.

We will show that the action in \er{eq:3d-eff} describes a gapped
phase.  In particular, all of the fermions obtain masses by non-zero VEV
of $a_3$ while the VEV preserves the $SU(N)$ gauge symmetry.  
We will indicate the role of $a_3$ after the compactification.  
In particular, we determine the surviving gauge symmetry $H\subseteq SU(N)$
after the compactification \footnote{The center symmetry of $SU(N)$ is explicitly broken by the fermions. For elaboration in the center symmetry, see, e.g., Ref. \cite{Kouno:2013mma}.}.

Let us consider a Wilson line around $S^1$,
\begin{align}
W={\cal P}\exp\left(i\int^{2\pi R}_0dx_3\ a_3\right),
\end{align}
which can generally be diagonalized as ${\rm diag.}(e^{i \xi_1}, e^{i
\xi_2}, \cdots, e^{i \xi_N})$, with $\sum_i^N \xi_i = 0$ (mod $2 \pi$).

The VEV of $a_3$ contributes to the mass of the KK modes of gauge
bosons as follows: 
\begin{align}
M_{n,ij}=\frac{1}{R}\left(n-\frac{\xi_i-\xi_j}{2\pi}\right).
\end{align}
If $\xi_i\neq\xi_j$ (mod $2\pi$), the KK mass spectrum breaks the
$SU(N)$ gauge symmetry\footnote{In other words, if the Wilson line $W$
and the generators of $SU(N)$ commute, the gauge symmetry is unbroken,
$H=SU(N)$.}~\cite{Hosotani:1983xw, Cossu:2013ora}.  

\begin{figure}[t]
\begin{center}
\includegraphics[width=7cm]{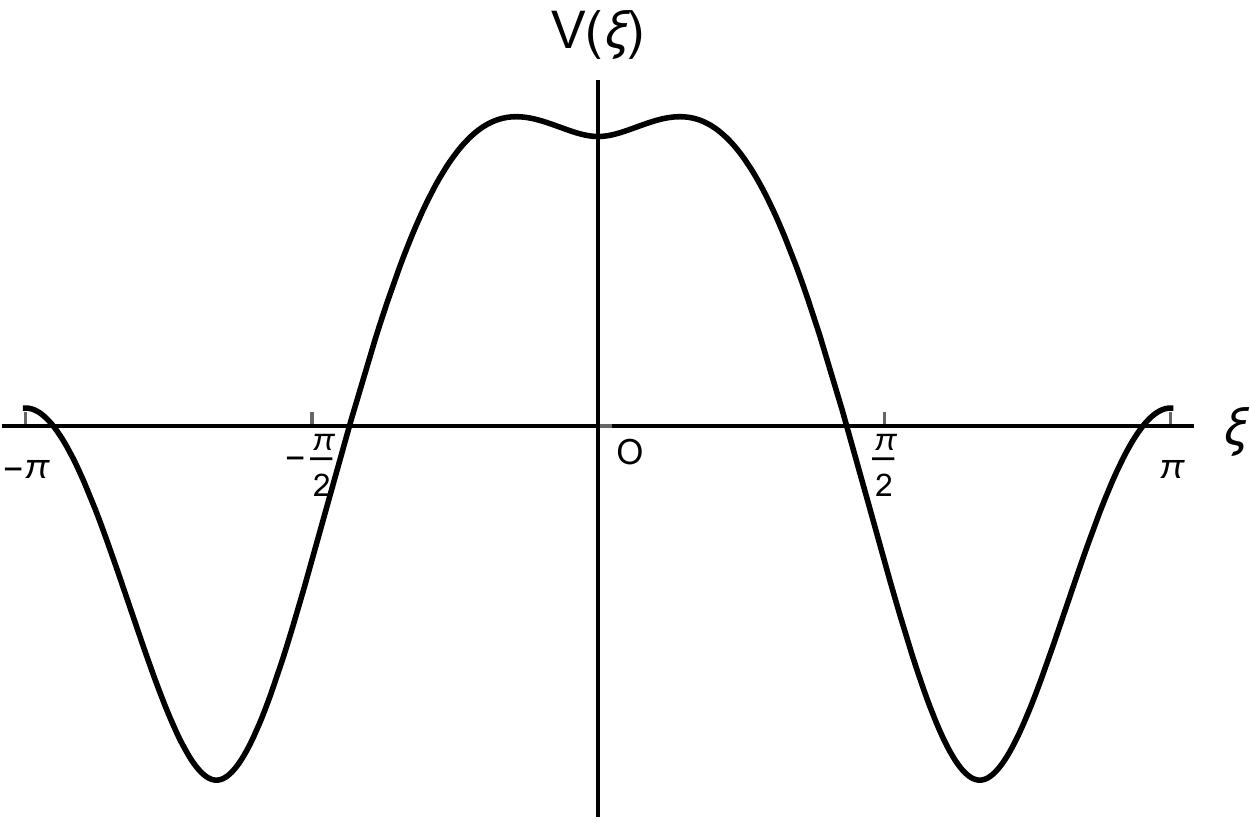} 
\hspace{0.5cm}
\includegraphics[width=7cm]{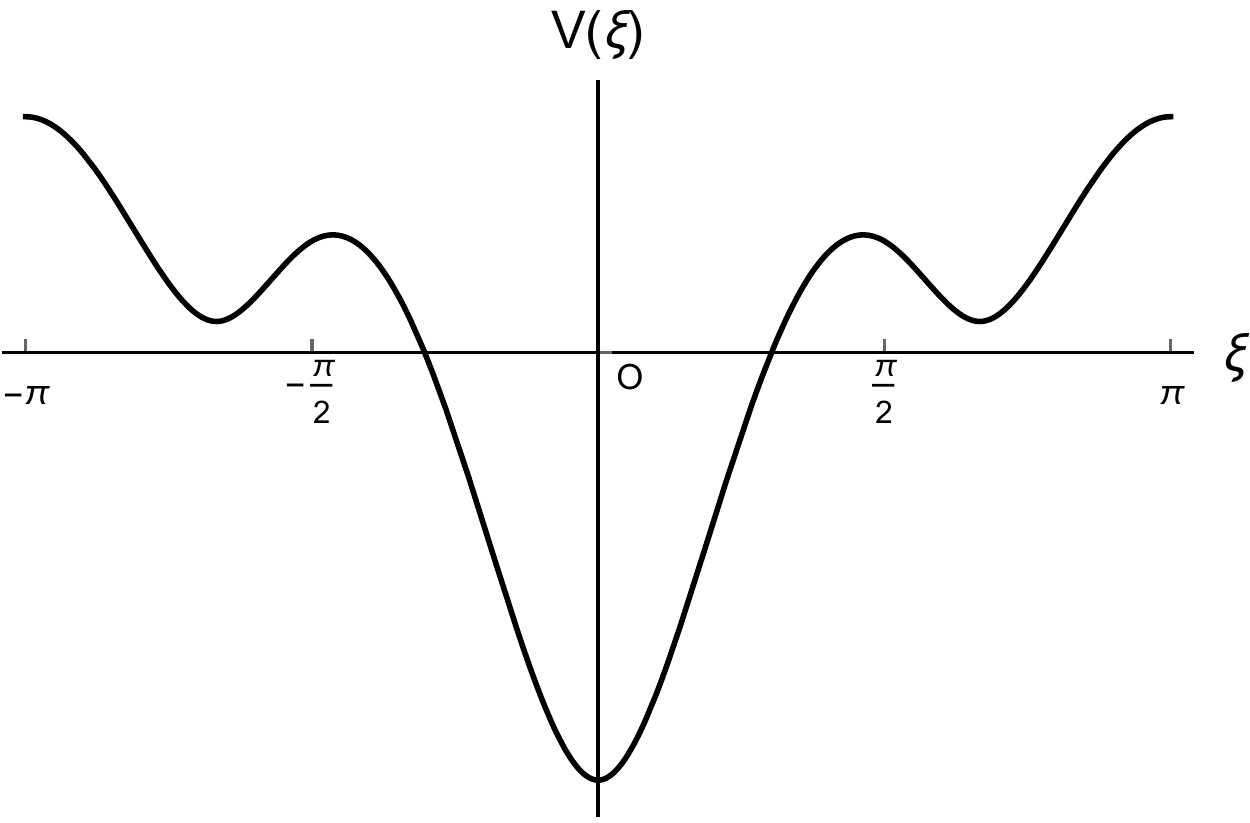} 
\end{center}
\caption{The shapes of the potential at $\nu = 0$ (left) and $\nu = \pi$ (right).}  
\label{fig:potential}
\end{figure}

The phases $\{\xi_i\}$ are determined dynamically by the potential
$V(a_3)$ which is generated at the one-loop level and it depends on the
boundary condition, $\nu$.  The one-loop effective potential as a
function of $\xi$'s is given by
\begin{align}
\label{eq:general_one_loop}
V(a_3)=&\frac{1}{8\pi^5 R^3}\Bigg[-2\sum^N_{i,j=1}\sum^\infty_{n=1}\frac{\cos n(\xi_i-\xi_j)}{n^4}\nonumber\\
&\qquad\qquad +2\sum^N_{i=1}\sum^\infty_{n=1}\frac{\cos n(-\xi_i+\nu+2\pi m_L)}{n^4}+(L\to R)\Bigg].
\end{align}
The potential is minimized when
\begin{align}
\label{eq:minimum_vev}
 a_3 = {1 \over 2 \pi R} {\rm diag.} (\xi, \xi, \cdots, \xi, - (N-1) \xi).
\end{align}
Inserting \eqref{eq:minimum_vev} to \eqref{eq:general_one_loop}, we obtain
\begin{align}
 V(a_3) = {1 \over 8 \pi^5 R^3}
\Bigg [
&
- 4 (N-1) \sum_{n=1}^\infty {\cos n N \xi \over n^4}
\nonumber \\
&
+ 2 N_f (N-1) \sum_{n=1}^\infty \left(
{\cos n (-\xi + \nu + 2 \pi m_L) \over n^4}
 + (L \to R)
\right)
\nonumber \\
&
+ 2 N_f \sum_{n=1}^\infty \left(
{\cos n ((N-1)\xi + \nu + 2 \pi m_L) \over n^4}
 + (L \to R)
\right)
\Bigg ].
\end{align}
The shapes of the potential of $\xi$ at $\nu = 0$ ($\nu
= \pi$) is shown in the left (right) panel of
Fig.~\ref{fig:potential} for $N=3$ and $N_f=2$.
In general, for $N_f < N$, there are $N$ minima at
$\xi = \pm 2 p \pi/N$, $(p=0,\cdots,[N/2])$. (For even $N$, $\xi =
\pi$ and $\xi = - \pi$ are equivalent.)
At $\nu = 0$, $\xi = 0$ is a local minimum, and the true minimum
is at $\xi = \pm (N-1)\pi/N$ for odd $N$ and $\xi = \pi$ for even $N$.
For $\nu = \pi$, the $\xi = 0$ point is the true vacuum.  

The potential has a symmetry $\nu \to \nu + 2 \pi n / N$, $n \in
{\mathbb Z}$, together with $\xi \to \xi - 2 \pi n / N$ that is the
reflection of the fact that the action of ${\mathbb Z}_N$ elements in
$U(1)_B$ is the same as that of the gauge group
$SU(N)$~\cite{Roberge:1986mm}.
For even $N$, $\nu = 0$ and $\xi = \pi$ is equivalent to $\nu = \pi$ and
$\xi = 0$. For odd $N$, they are not equivalent. The $\nu = \pi$ point is
equivalent to $\nu = \pi / N$ by an appropriate shift of $\xi$.
There is a first order transition in between $\nu = 0$ and $\nu = \pi
/N$.

In all the $N$ minima, the $SU(N)$ gauge symmetry is unbroken as the
Wilson loop along the $x_3$ direction is a phase times the unit
matrix. The fermion masses for $\psi_n$ and $\tilde \psi_n$ are,
respectively,
\begin{align}
 m_n^{(\psi)} = - {n \over R} - {\nu \over 2 \pi R} - {m_L \over R} + {\xi \over 2 \pi R}, \quad
 m_n^{(\tilde \psi)} = {n \over R} + {\nu \over 2 \pi R} + {m_R \over R} - {\xi \over 2 \pi R}.
\end{align}
By following the global minimum of the potential, in the entire region of
$\nu$, the fermion masses are non-vanishing.
Therefore, the low energy limit of the 4d QCD on $M^3 \times S^1$ is
$SU(N)$ pure gauge theories on $M^3$ for a small radius. There is a mass
gap, but the low energy limit can be a topological field theory.
For $m_R = 1$ and $m_L = 0$, the fermion masses for $\tilde \psi_n$ are
shifted by $1/R$.
The shift changes the sign of $N_f$ fermion masses from negative to
positive. Therefore, the low energy theory obtains the CS level $N_f$.
($N_f/2$ to integrate in the negative ones, and another $N_f/2$ to
integrate out the positive ones.)

The CS level is consistent with the result in the basis of $k=N_f$. The
$\theta$ term can be expressed as
\begin{align}
{1 \over 8 \pi^2} \int_{M^3 \times S^1} {\theta } {\rm Tr} (f f)
&=
{1 \over 8 \pi^2 }\int_{M^3 \times S^1} 
{\rm Tr} \left(
a da + {2 \over 3} a^3 \right) d \theta, \mod 2 \pi.
\end{align}
Again, it is important that the integral on
$S^1$ is properly defined~\cite{Cordova:2019jnf}.
Since $d \theta$ is single valued, one can naively use the right-hand
side of the integral over $S^1$, which reduces to the CS term with the
level $k$ for the lowest KK mode of the gauge field.

In summary, the low energy limit of QCD$_4$ on $M^3 \times S^1$ with
Eq.~\eqref{eq:param} for a small radius, $\Lambda_4 R \ll 1$, is the
topological field theory, $SU(N)_{N_f}$, that has the dual description
by $U(N_f)_{-N}$.

\subsection{Large radius, $\Lambda_4 R \gg 1$}
One can also analyze the low energy limit of QCD$_4$ on $M^3 \times S^1$
for a large radius as we know that the low energy theory is described by
pions. We discuss the effect of $\bar \theta (x_3)$ in the low energy
effective theory.
In order to see the $\bar \theta (x_3)$ dependence of the theory, one
needs to introduce $\eta'$ meson
in addition to the massless pions.  Note that a similar topic where
there is a phase transition as a function of the gradient of
$\bar\theta$ is established in Ref.~\cite{Gaiotto:2017tne}.

The effective theory is given in terms of the $N_f \times N_f$ unitary
matrix 
\begin{equation}
 U = e^{i \pi^a T^a + i \eta'}.
\label{eq:meson}
\end{equation}
We implicitly assume here
that $\eta'$ behaves as a Nambu-Goldstone boson,
which is true in the large $N$ limit.
Hence, the results are formally correct to the leading order 
in the $1/N$ expansion. 
The field $U$ transforms as
\begin{align}
 U \to g_L^{-1} U g_R,
\label{eq:Utransf}
\end{align}
under the $g_L \in SU(N_f)_L$ and $g_R \in SU(N_f)_R$ chiral
transformations. Under the axial $U(1)_A$, it transforms as $U \to e^{2
i \beta} U$.
The effective action is given by
\begin{align}
 S_{\rm eff} = &
\int_{M^3 \times S^1} d^4 x 
\nonumber \\
&
\times \bigg [
{f_\pi^2 } {\tr} 
| \partial_\mu U |^2
- {m_{\eta'}^2 f_\pi^2 \over N_f} \left|
\log (e^{- i \bar \theta } \det U)
\right|^2
+ \cdots
\bigg].
\label{eq:eff}
\end{align}
Here, we have added the mass term of $\eta'$, which is generated at 
the order of $1/N$.
The effect of the boundary condition, $\nu$, can be taken into account
by introducing a background gauge field for the baryon number, $(\nu / 2
\pi R) \bar \Psi \gamma^3 \Psi$. The effect of this term appears in the
WZ terms. The parameter $\nu$ couples to the topological current, i.e.,
the Skyrmions.

In what follows we show that ${\eta'}$ develops a winding number due to the
winding of $\bar\theta$.  Here we consider the case $\bar k =N_f$ as
before and discuss the case with smaller values of $\bar k$ later.  In
this case we obtain the same WZ term as in the three dimensional
$U(N_f)_{-N}$ theory by integrating out ${\eta'}$.
At the linearized order, the equation of motion for
${\eta'}$ is
\begin{align}
 {\partial^2 \over \partial x_3^2}{\eta'}
= m_{\eta'}^2 
\left( {\eta'} - {\bar \theta \over N_f} \right).
\end{align}
Here, we have used the parameterization of $U$ in \er{eq:meson},
and neglected the derivatives along the $x_{0,1,2}$ directions 
which are irrelevant for the discussion here.
For the periodicity of $\bar \theta$, the potential for ${\eta'}$ has $N_f$
domains where the ${\eta'} - \bar \theta / N_f$ is minimized at ${2 n \pi
/ N_f}$ with $n$ integers.
Transition between two other domains requires a treatment beyond this
effective theory.
%
If we require that
the shape of the function $\bar \theta$
is not very rapid so that the effective theory can be used, ${\eta'}$
should stay in one of the domains to minimize the energy, 
which means that $\eta'$ develops a winding number under 
Eq.~\eqref{eq:param}:
\begin{align}
 {\eta'} (x_3 + 2 \pi R) = {\eta'} (x_3) + 2 \pi.
\end{align}
This is consistent with the $S^1$ compactification.
%
Note that the non-zero winding number of $\eta'$ does not imply the jump
of domains.
We are treating ${\eta'}$ as a heavy field and we are
working within the effective theory. Also, the argument of the winding
of ${\eta'}$ depends on the basis. We can eliminate $\bar \theta$ by the
field redefinition of ${\eta'}$.  
In this basis $\eta'$ does not develop a winding number,
 while we obtain the same physics at low energy, which we will
discuss in the next paragraph after introducing the WZ term in
\er{eq:WZ}.

The three dimensional low energy effective theory is the non-linear
sigma model with the coset in Eq.~\eqref{eq:coset}, but there are
effects from the ${\eta'}$ winding.
By turning on the external gauge field, $A$, which couple to the vector
current as we discussed in the previous section, a part of the WZ term,
\begin{align}
 S_{\rm WZ} = - {N \over 8 \pi^2} \int_{M^3 \times S^1}  {\rm Tr}
\left(
A d A + {2 \over 3} A^3
\right) d {\eta'},
\label{eq:WZ}
\end{align}
reduces to Eq.~\eqref{eq:WZext} where the minus sign is chosen, i.e., it
is the same as the one obtained from $U(N_f)_{-N}$ theory. It is
interesting that the external magnetic field carries baryon number in
this ${\eta'}$ winding background. See Appendix~\ref{sec:baryon} for the
relation between the profile of ${\eta'}$ and the baryons.
The parameter set in Eq.~\eqref{eq:thetazero} corresponds to the basis
where ${\eta'} - \bar \theta / N_f$ is redefined to be ${\eta'}$. In that
bases, there is no winding of ${\eta'}$, but the same WZ term appears by
shifting ${\eta'}$ in Eq.~\eqref{eq:WZ}.
One can confirm the consistency of the appearance of the WZ term by
comparing the $\bar \theta \to AA$ amplitude to that in QCD$_4$.

In summary, we find that the low energy limit of QCD$_4$ with
Eq.~\eqref{eq:param} is a topological field theory, $SU(N)_{N_f}$, for
small $\Lambda_4 R$ and a non-linear sigma model with the WZ term for a
large radius, $\Lambda_4 R \gg 1$.
There must be a phase transition between these two extreme regions.
It is interesting to find that the two limits are the same as the
conjectured limit of the three dimensional $SU(N)_0$ theory with $2N_f$
fermions with large and small fermion masses.
It is therefore possible to anticipate that the phase transition between
a large and a small radius is described by the critical point of
$SU(N)_0$ QCD$_3$ with $2N_f$ fermions with the explicit $U(2N_f)$
symmetry breaking terms.
The phase transition can also be consistent with the description by the
three dimensional $U(N_f)_{-N}$ theory with $2N_f$ scalar fields.

For $|\bar k| < N_f$ instead of Eq.~\eqref{eq:param}, the ${\eta'}$ winding
should be accompanied with the winding of pions. 
In order to satisfy the boundary condition, $U(x_3 + 2 \pi R) = U(x_3)$,
the winding ${\eta'} (x_3 + 2 \pi R) = {\eta'} (x_3) + {2 \pi \bar k / N_f}$
should be accompanied by
\begin{align}
 e^{i \pi^a T^a (x_3 + 2 \pi R)} = e^{- 2 \pi i \bar k / N_f} e^{i \pi^a T^a (x_3)},
\end{align}
that can be realized as a configuration of $\pi^a$ as the phase factor
is an element of $SU(N_f)$.
For example, for $\bar k=1$, one of the diagonal components of $U$
acquires a winding by $2 \pi$, and the non-trivial WZ term appears only
for that component of the external gauge fields.
In general for $|\bar k| \le N_f$, the same effective three dimensional
theory can be obtained by the theories of pions interacting with a
Higgsed $U(|\bar k|)_{\mp N}$ gauge group.
This part matches the conjectured dualities in three dimensions between
$SU(N)_{\bar k-N_f}$ with $2N_f$ fermions and $U(\bar k)_{-N}$ with
$2N_f$ scalars.

\section{Hadrons near the critical point}
\label{sec:hadron}

Here we proceed to speculations on the possible behavior of the vector
mesons based on the discussion in the previous section.
As we discussed, QCD$_4$ in the background of Eq.~\eqref{eq:param}
provides us with the same low energy theories of QCD$_3$ both in the
broken and unbroken phases of the global symmetry once an explicit
breaking term of the $U(2N_f)$ symmetry is added. The phase transition
in three dimensions has a dual picture by the $U(N_f)_{-N}$ gauge
theory.
Below, 
we will consider the possibility that the dual picture also describes
the QCD$_4$ near the phase transition.

Indeed, it is interesting that the extension of the chiral Lagrangian to
a $U(N_f)$ gauge theory is known to give a great success to describe the
phenomenology of the vector mesons, 
$\rho$ and $\omega$~\cite{Bando:1984ej,Bando:1985rf,Bando:1987br}.
Therefore, the vector mesons are the natural candidates for the gauge
bosons of the $U(N_f)$ dual theory.
If such an interpretation is true, in the phase where chiral symmetry is
restored, the $\rho$ and $\omega$ mesons are in the topological phase
rather than in the Higgs phase under the background in
Eq.~\eqref{eq:param}.

The possible behavior of the hadrons as the function of the radius is as
follows.
Starting from a large radius where the hadrons describe physics
effectively, as the radius approaches to the critical point, $R_* \sim
1/\Lambda_4$, the lowest mode of hadrons start to form a $U(N_f)_{-N}$
gauge theory in the Higgs phase. The members are pions, $\rho$, $\omega$
and other scalar mesons.
At the critical point, the chiral symmetry is restored and the $\rho$
and $\omega$ mesons get into the topological phase described by the
$U(N_f)_{-N}$ theory, that is dual to $SU(N)_{N_f}$.
As further decreasing the radius, the picture of weakly interacting
gluons and quarks makes sense at the energy scale between $\Lambda_3$
and $1/R$. All the fermions as well as KK modes of gluons decouple below
$1/R$. Although the description in terms of hadrons gets ineffective in
this energy region, the low energy limit of the theory stays the same.

Since $U(N_f)$ gauge group is spontaneously broken, there are vortex
configurations in three dimensions made of $\rho$ and $\omega$, which
carry magnetic and electric charges of $U(1)^{N_f}\ (\subset
U(N_f))$~\cite{Paul:1986ix}.
The electric charge is a consequence of the CS term. In the color-flavor
locked phase, the electric charge is identified as the baryon number. The
vortex with the unit magnetic charge has $B=1$.
We will discuss how this vortex configuration extends to the $S^1$
direction in the next section.

\section{Holographic model}
\label{sec:holographic}

The vector mesons as gauge bosons are nicely described by the
holographic QCD where the gauge bosons are propagating into an extra
dimension~\cite{Sakai:2004cn, Erlich:2005qh, DaRold:2005mxj}. Based on
the holographic model, we here try to find a dual model of QCD$_4$ which
reproduces the story in the previous section.
For the duality to work, we need the same low energy limits both in the
broken and unbroken phase of the chiral symmetry.
In particular, one needs to arrange the theory such that there is
an unbroken gauge group, $U(N_f)_{-N}$, in the symmetric phase.
The chiral symmetry breaking is described by the VEVs of scalar fields
which simultaneously make the gauge group Higgsed.

Of course, we are not aware if the phase transition is smooth enough
that such an effective description exists. 
We here assume that the phase transition is the
second order, and look for a dual theory.
\footnote{In this paper, we consider the massless fermions with 
the periodic boundary condition along $x_3$ direction, 
and the number of color is assumed to be greater than the number of flavor, $N > N_f$.
If we instead impose the anti-periodic boundary condition,
the order of the transition would be changed (see e.g., Ref.~\cite{Cuteri:2017zcb}).}
In this sense, this is a construction of the Nambu-Jona-Lasino model or
the Ginzburg-Landau model while taking into account the consistency with
topology.
For $\bar k=0$ it was a trivial task since the low energy limit of symmetric
phase is trivially gapped. But for $\bar k \neq 0$, we need some gauge theory
to survive to match the topological field theory.
In this section, we mainly consider the background in Eq. \eqref{eq:param}.
We will comment other backgrounds in the last part of this section.

The holographic QCD describes the vector mesons and pions as gauge
fields propagating into the fifth dimension, ${b}_{L,R}$.
The gauge group is $U(N_f)_L \times U(N_f)_R$ which is broken down to
$U(N_f)_{L+R}$ somewhere in the extra dimension.
The five-dimensional space has a boundary. The boundary conditions are
taken to be
\begin{align}
 {b}^\mu_{L,R} \Big|_{\rm boundary} = A_{L,R}^\mu, \quad (\mu = 0,1,2,3),
\label{eq:boundary}
\end{align}
where the right-hand side is external gauge fields which couple to
chiral currents. The pions appear as the extra dimensional component of
${b}_{L}^4 - {b}_R^4 \sim \partial^4 \pi$. The boundary
condition makes the gauge bosons massive, and the lightest modes are
identified as the $\rho$ and $\omega$ mesons.
The WZ terms can be reproduced by
\begin{align}
 S_{\rm CS} = - {N \over 24 \pi^2} \int_{X_5} 
\left(
\omega_5 ({b}_L)
- \omega_5 ({b}_R)
\right).
\label{eq:CS-5d}
\end{align}
See Ref.~\cite{Domenech:2010aq} for details. The local chiral
transformation shifts the external gauge fields and modifies the
boundary conditions, that can be absorbed by the gauge transformation of
the bulk gauge fields, and that in turn provides a boundary term from
the gauge transformation of the CS term. This procedure results in the
WZ terms on the boundary written in terms of the pions and the external
gauge fields. They are necessary to match the 't~Hooft anomaly in
QCD$_4$.

\begin{figure}[t]
\begin{center}
 \includegraphics[width=7cm]{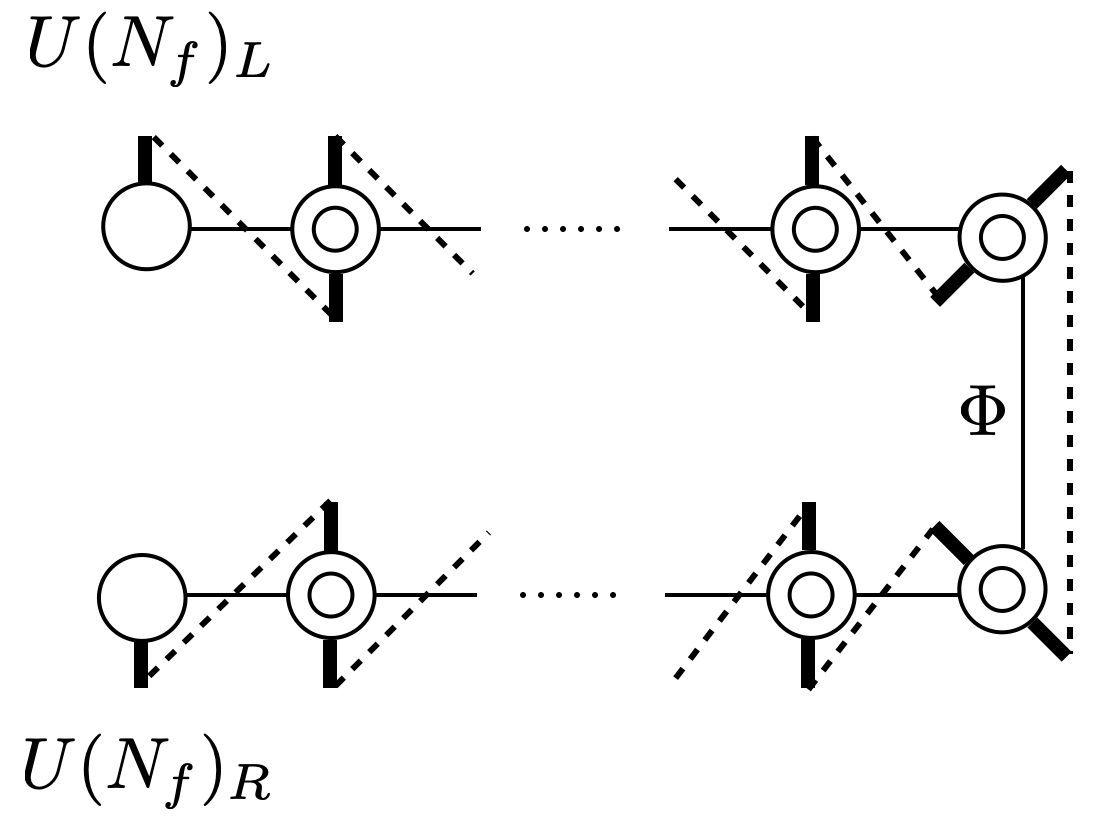} \hspace{5mm}
 \includegraphics[width=7cm]{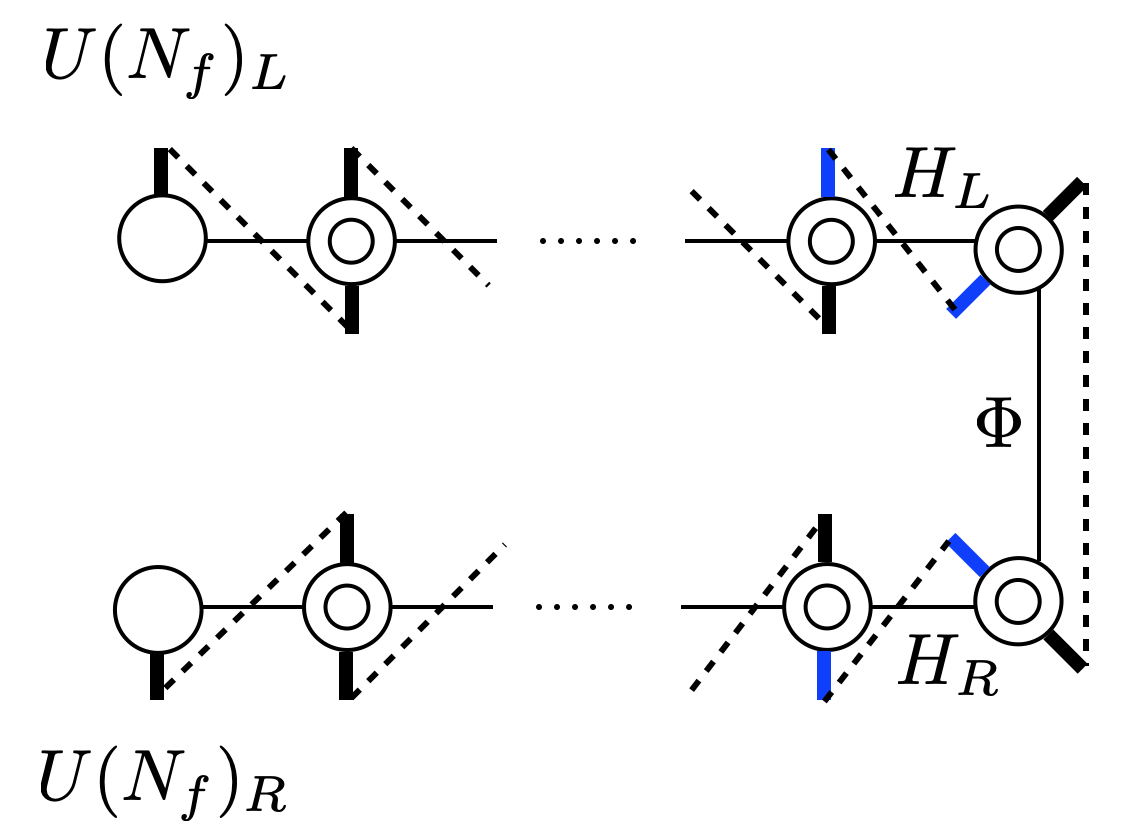}
\end{center}
\caption{The quiver diagram for the holographic QCD (left) and the
 linearized model (right).}  \label{fig:quiver}
\end{figure}

The deconstructed version of the model can be built as in the left panel
of Fig.~\ref{fig:quiver}~\cite{Son:2003et}. The explicit form of the
Lagrangian is given in Appendix~\ref{sec:quiver}.
The most left sites with open circles are the boundary. No gauge fields
are living there.
At other sites with double circles, there are $U(N_f)$ gauge fields
which are all Higgsed by eating the link fields.
By the most right link, denoted $\Phi$, the $U(N_f)_L \times U(N_f)_R$
symmetry is broken down to $U(N_f)_{L+R}$.
Since the number of links are larger by one than the number of double
circles, there are $N_f^2$ Nambu-Goldstone bosons left uneaten. They are
identified as the pions and ${\eta'}$.
The thick lines in each site represent $N$ chiral fermions. The dashed
line between fermions represents the mass terms. They are all massive,
but necessary to reproduce the CS term in the five dimensional theory.
Since the fermions are always massive even in the chiral symmetric phase
as we see later, we can think of fermions as auxiliary degrees of
freedom.
We also gauge the ${\mathbb Z}_N$ subgroup of $U(1)_B$ which transforms
fermions. The gauging is necessary to match the global symmetry $U(1)_B
/ {\mathbb Z}_N$ in the original QCD.
The gauging requires that one should sum up all the sectors with
boundary conditions of fermions twisted by elements of ${\mathbb Z}_N$ in
the $x_3$ direction.

Now let us turn on the external gauge fields. The external gauge fields
couple to fermions as
\begin{align}
 A_{L}^\mu 
\left( \bar q_L \gamma_\mu q_L + \cdots \right)
 +  (L \leftrightarrow R).
\end{align}
The fermions in the upper and lower wings couple to $A_{L}$
and $A_{R}$, respectively. The mass terms of fermions except
for the one with the link $\Phi$ do not break the global symmetry,
$U(N_f)_L \times U(N_f)_R$.
The combination of $A_{L}+A_{R}$ couples to the conserved vector
current.
By integrating out the massive fermions, one obtains the correct WZ
terms.
At this stage, we have not included the mass term of ${\eta'}$. One can
introduce it by writing a mass term to break the $U(1)_{L-R}$ gauge
symmetry for $\Phi$, such as
\begin{align}
 |\log \det \Phi|^2.
\end{align}
The axial $U(1)$ is now explicitly broken, and the ${\eta'}$ obtains the
mass.
In the background of Eq.~\eqref{eq:qcd}, we have a term
\begin{align}
 |\log e^{- i \bar \theta} \det (\Phi) |^2.
\end{align}
This term cause the winding of the trace part of $\Phi$.
The winding gives important effects through the WZ term as we discussed
before.
In addition to the WZ terms among external fields and Nambu-Goldstone
modes, we also have
\begin{align}
 S_{\rm WZ} = - {N \over 8 \pi^2} \int_{M^3 \times S^1}  {\rm Tr}
\left(
{b} d {b} + {2 \over 3} {b}^3
\right) d (- i \log \det \Phi)  / N_f,
\label{eq:CSbulk}
\end{align}
where ${b}$ is a gauge field of $U(N_f)_{L+R}$ part of the right most
gauge sites. They are massive modes which correspond to the $\rho$ and
$\omega$ mesons.

One can modify the model by introducing Higgs fields, $H_L$ and $H_R$,
as in the right panel of Fig.~\ref{fig:quiver}. Two of the links are
replaced by the Higgs fields.
When the VEV of $\langle H_{L,R} \rangle$, proportional to the unit
matrix, is large, one can identify the sites on the both sides of
$H_{L,R}$, and it comes back to the model in the left panel.
The location of the links to be replaced with $H_{L,R}$ can be anywhere
in the wings.

For small $\langle H_{L,R} \rangle$, the vector part of the gauge bosons
in the most right sites, i.e., $\rho$ and $\omega$, gets light, and for
$\langle H_{L,R} \rangle = 0$, the gauge bosons as well as the fermions
(marked as blue lines) become ``massless.'' They obtain masses when the
$x_3$ direction is compactified on $S^1$.
All the link fields are eaten by
the gauge fields, and thus massless pions disappear. Therefore, the
chiral symmetry is now recovered in this phase.
In this way, the model interpolates the chiral Lagrangian and the linear
sigma model by changing the sizes of $\langle H_{L,R} \rangle$.

In the phase of $\langle H_{L,R} \rangle = 0$, we have the term in
Eq.~\eqref{eq:CSbulk} but now ${b}$ represents the massless gauge boson.
For the background with $\bar k = N_f$, the winding of $\Phi$ gives the
winding $\theta$ term for vector mesons.

Let us consider the vacuum of the theory after the compactification of
the $x_3$ direction by the one-loop effective potential near the phase
transition point, $\langle H_{L,R}\rangle=0$.  In this region, the
massless degrees of freedom are the $U(N_f)$ gauge field and $4N$ chiral
fermions. The scalar fields $H_{L,R}$ can be light, but the contribution
of the scalars has the same shape as the fermions with the opposite sign
and thus for $N_f < N$, the effects can be ignored.
The half of $4N$ massless fermions are charged under $U(N_f)$ while the
rest are neutral. We first discuss the contribution to the effective
potential of $b_3$ from the $2N$ chiral fermions.
We take an ansatz of the VEV of $b_3$ to be
\begin{equation}
b_3=\frac{1}{2\pi R}{\rm diag.}(\tilde{\xi}_1,\tilde{\xi}_2,\dots,\tilde{\xi}_{N_f}),
\end{equation}
where $\sum_{i=1}^{N_f}\tilde{\xi}_i=0\ ({\rm mod}\ 2\pi)$.
The boundary condition of the fermions is
\begin{equation}
q(x_3+2\pi R)=e^{i\nu} q(x_3),
\end{equation}
where $0\leq\nu<2\pi$.
The KK spectra of the gauge fields and the fermions are, respectively,
\begin{equation}
M_{ij,n}^2=\frac{1}{R^2}\left(n-\frac{\tilde{\xi}_i-\tilde{\xi}_j}{2\pi}\right)^2,
\end{equation}
\begin{equation}
m_{i,n}^2=\frac{1}{R^2}\left(n+\frac{\nu-\tilde{\xi}_i}{2\pi}\right)^2.
\end{equation}
By using these, the one-loop effective potential is given by
\begin{equation}
V(b_3)=\frac{1}{4\pi^5 R^3}\left(-\sum_{i,j=1}^{N_f}\sum_{n=1}^{\infty}\frac{\cos(n(\tilde{\xi}_i-\tilde{\xi}_j))}{n^4}+2N\sum_{i=1}^{N_f}\sum^\infty_{n=1}\frac{\cos(n(\nu-\tilde{\xi}_i))}{n^4}\right).
\end{equation}
The first term in the right hand side is minimized when $\tilde{\xi}_1=\tilde{\xi}_2=\cdots=\tilde{\xi}_{N_f}=\tilde{\xi}$.
Furthermore, the second term is minimized at $\tilde{\xi}=\nu-\pi$.
Therefore, for $N_f < N$, the lowest minimum is at $\int_{S^1} {b} =
(\pi - \nu) \cdot {\bf 1}$, which gives the anti-periodic boundary
conditions for the fermions. The $U(N_f)$ group is unbroken at the
minimum, and the $2N$ fermions get massive.

Next, we consider the one-loop effects of the gauged $\mathbb{Z}_N$
which is the subgroup of $U(1)_B$ symmetry.  The effective potential is
given by
\begin{equation}
V(\xi')=\frac{N}{2\pi^5 R^3}\sum^\infty_{n=1}\frac{\cos(n(\nu-\xi'))}{n^4},
\end{equation}
where $\xi'=2\pi m/N\ (m\in\mathbb{Z})$ is the VEV of the gauged
$\mathbb{Z}_N$ field. The different VEV corresponds to the sector of
different boundary conditions twisted by the ${\mathbb Z}_N$ elements.
Since ${\mathbb Z}_N$ is gauged, we are summing up all the values of
$\xi'$ in the path integral.
The free energy is minimized at $m=N(\nu-\pi)/(2\pi)$, which means the
path integral is dominated by this sector.
The anti-periodic boundary condition is chosen for even $N$, and the
sector which is the closest to the anti-periodic one is chosen for odd
$N$.
Therefore, all the fermions decouple and the effective three dimensional
theory is the $U(N_f)_{-N}$ CS theory where the CS level stems from the
winding of $\Phi$. This theory is dual to $SU(N)_{N_f}$ that is the low
energy limit of QCD$_4$ for a small $S^1$ radius.

It is important that the axial $U(1)$ is kept broken in the phase of
$\langle H_{L,R} \rangle = 0$ where chiral symmetry $SU(N_f)_L \times
SU(N_f)_R$ is unbroken. The explicit breaking is important to obtain the
correct CS level via the winding of $\Phi$.
Therefore, for the scenario to work the axial $U(1)$ should be broken
during the chiral phase transition.
\footnote{The breaking of the axial $U(1)$ symmetry 
implies that topological susceptibility does not vanish during 
the chiral phase transition
(see e.g., Refs. \cite{Borsanyi:2016ksw,Tomiya:2016jwr} 
for discussions following lattice simulations).}

The baryon number as the topological charge in the $U(N_f)_{-N}$ theory
can be seen in this model.
The baryon number in this model is identified as the trace part of the
external $U(N_f)_{L+R}$. 
The unbroken baryon number is rearranged to be a sum of all the $U(1)$
part of each site, and thus the fermions which are integrated out to
obtain Eq.~\eqref{eq:CSbulk} are then charged under the baryon number.
Therefore, by turning on the background gauge field, $B$, for the baryon
number we obtain a term in the three dimensional effective theory:
\begin{align}
 S_{\rm baryon} = - {1 \over 4 \pi} \int_{M^3} B \, {\rm Tr} (d {b}),
\end{align}
under the winding of $\Phi$. 
Here, we have integrated $\Phi $ in \er{eq:CSbulk}
in the presence of the winding number along the $x_3$ direction.

As in the three dimensional effective picture, there are vortex
configurations made of $\rho$ and $\omega$. Under the winding of $\Phi$,
there is a winding $\theta$ term for the $U(N_f)$ gauge group from
Eq.~\eqref{eq:CSbulk}. In the presence of this $\theta$ term (which we
call it $\tilde \theta$), the Abrikosov-Nielsen-Olesen (ANO) vortex
string~\cite{Abrikosov:1956sx, Nielsen:1973cs} which goes around $S^1$
cannot be connected since the magnetic flux obtains the electric charge
as it goes around the $S^1$ direction by the Witten
effect~\cite{Witten:1979ey}
(see appendix \ref{sec:baryon} for the detail).
In order to have a stable string loop which goes around $S^1$, one needs
to have some non-trivial configuration which carries the electric
charge, {\it i.e.,} the baryon number.

In the background with a constant $d \tilde \theta$, one can look for a
static field configuration which is $x_3$ independent. The field
equations are then the same as the CS case, and thus one finds the
solution with a finite energy. The baryon number, $B=1$, is indeed
carried by the string through the electric charge of this solution. For
a general background, this configuration will be relaxed to a solution
of the field equations with a finite energy.
We call it the $B=1$ string.

Another possibility is to unwind $\Phi$, {\it i.e.,} ${\eta'}$, by forming
a Hall droplet described in Ref.~\cite{Komargodski:2018odf}.
The droplet is a configuration of the ${\eta'}$ field. It is a sheet with a
boundary, and the value of ${\eta'}$ changes by $\pm 2 \pi$ when we go
across the sheet.
The ANO $\rho-\omega$ string can be connected when it goes across the
sheet as the Witten effect is cancelled by the change of $\tilde
\theta$.
The net baryon number of this configuration is $B=0$, and thus we call
it the $B=0$ string. This string is penetrating the Hall droplet. The
droplet cannot shrink to nothing, as there is no $B=0$ string in the
background without the droplet.
The $B=1$ string discussed above cannot smoothly
deform into this string due to the different baryon number.
Interestingly, it is proposed in Ref.~\cite{Komargodski:2018odf} that
the droplet has an excitation of the edge mode with $B=1$ and that
object is identified as the baryon such as $\Delta^{++} \sim uuu$. Therefore,
the $B=1$ string can deform into a $B=0$ string together with the
excitation of the edge mode of the droplet with $B=1$.
Since there is no such stable string configuration in full QCD, we
expect that there are monopoles to cut the $B=0$ string. A pair of a
monopole and an anti-monopole can cut and eliminate the string. Now the
$B=1$ string can decay into a baryon via the deformation into a
$B=0$ string and a $B=1$ droplet and then the $B=0$ string part is
eliminated.
This is a good candidate of the fate of the $B=1$ string.

So far, we have assumed the background in Eq. \eqref{eq:param}.
The phase transition by the VEV of $H_{L,R}$ can be extended to the case
of $0 \le \bar k < N_f$.
In that case, the unbroken gauge group is $U(\bar k)_{-N} \times U(N_f -
\bar k)_0$. In order for the theory to have the same low energy limit as
QCD, the $U(N_f - \bar k)_0$ factor should decouple. 
The $SU(N_f - \bar k)_0$ part exhibits a mass gap by the confinement. 
The $U(1)$ part also confines by instantons.
As we discussed in the previous section, we need monopoles (in four
dimensions) to cut the stable $\rho$ and $\omega$ strings. The monopole
configurations in $(012)$ directions are instantons in three dimensions,
and the path integral including such instantons causes the confinement
of the $U(1)$ factor~\cite{Polyakov:1975rs, Polyakov:1976fu}.
The presence of the monopole does not make the $U(\bar k)_{-N}$ part 
confine as the gauge bosons have a mass term from the CS
term~\cite{Affleck:1989qf}.
Therefore, we obtain the correct low energy limit, $U(\bar k)_{-N}$
theory.

There is another possibility that the $U(N_f - \bar k)$ part of the VEV
is kept non-vanishing for $H_{L,R}$ while the chiral symmetry is
restored by cutting the $U(N_f - \bar k)$ part the link $\Phi$.
The $U(N_f - \bar k)$ gauge group is kept in the Higgs phase, and the
only $U(\bar k)_{-N}$ part remains at low energy.

What happens for $\rho$ and $\omega$ mesons is qualitatively different
in the above two cases.
When the $S^1$ radius is large, they are vector mesons which we are
familiar with.
As the radius approaches to the critical point, they behave as the gauge
bosons in the Higgs phase. In particular, the meta-stable vortex strings
made of $\rho$ and $\omega$ appear.
Beyond the critical radius, the chiral symmetry is restored, and the
$U(\bar k)$ part of them goes into the topological phase, whereas the
$U(N_f - \bar k)$ part goes into either the confining phase or remains
in the Higgs phase. The rest of them stays in the Higgs phase.

The model described here can be viewed as the Nambu-Jona-Lasino model
for chiral symmetry breaking under the background of the imaginary
chiral chemical potential (once we take the compactification direction
to be the time direction in the Euclidean space.)
The ${\mathbb Z}_N$ twisting boundary condition can be naturally
identified as the VEV of the Polyakov loop. For a small radius (high
temperature), $\langle H_{L,R} \rangle$ is vanishing, and thus the
fermions choose a particular boundary condition by minimizing the free
energy. This corresponds to the non-vanishing VEV of the Polyakov loop,
and thus describing the deconfined phase. On the other hand, for a large
radius where $\langle H_{L,R} \rangle$ is large, the fermions decouple,
and all the boundary conditions equally contribute to the path
integral. This corresponds to vanishing VEV for the Polyakov loop, and
thus the quarks are confined.

\section{Finite temperature QCD}
\label{sec:finiteT}

We discussed a somewhat exotic scenario for the chiral phase transition
which happens at some critical radius, $R_*$.
Let us apply our results for the finite temperature QCD.
Here, we mainly consider our results for a general $\bar{k}$ 
in \er{eq:combi} 
rather than $\bar{k} = N_f$ in \er{eq:param}.
In particular, we mainly discuss the $\bar{k} =0$ case,
where the physical phase, $\bar\theta$, is absent. 
For $\bar k = 0$ 
with the anti-periodic boundary condition for quarks,
$\nu = \pi$, one can think of this system as the finite temperature QCD
by considering the Euclidean metric. At some critical temperature, 
$T_* =1/R_*$, the chiral phase transition happens.

In the model we discussed, there is a consistent scenario where $U(\bar
k)_{-N} \times U(N_f - \bar k)_0$ theory remains in the infrared and
$U(N_f - \bar k)_0$ factor confines due to instantons. For $\bar k = 0$,
the confining $U(N_f)_0$ gauge field is the $\rho$ and $\omega$ mesons.

There are other possibilities as we discussed already. There is also a
possibility that $U(N_f)$ gauge theory is not a good picture at all.
For example, the gauged Nambu-Jona-Lasino model in
Ref.~\cite{Kondo:1992sq} can give $SU(N)_{\bar k}$ factor in the chiral
symmetric phase. The model is simply adding to QCD a scalar field, $X$,
which transforms as $(N_f, \overline{N_f})$ under the $U(N_f)_L \times
U(N_f)_R$ chiral symmetry, and coupling it to quarks as $\bar q X
q$. This model gives the correct non-linear sigma model in the broken
phase where the scalar field has a VEV, while it reduces to $SU(N)_{\bar
k}$ theory in the symmetric phase. The $\rho$ and $\omega$ mesons do not
appear as the field to describe the phase transition phenomena.

The question of which picture is the most appropriate near the chiral
phase transition should be able to be tested by the lattice
simulations. By looking at the behavior of the two point functions of
the vector currents, one may check if the $\rho$ and the $\omega$ mesons
get ``massless.''
However, it does not mean that the screening masses of 
the $\rho$ and $\omega$ mesons vanish.
Although they have no mass term in the four dimensional Lagrangian, they
have thermal masses and also masses from instantons (monopoles in four
dimensions) in the actual spectrum. We will leave the study of these
effects as well as that of actual methods in the lattice QCD to
distinguish the scenarios.
For the lattice simulations of the screening masses of them,
see, e.g., 
Refs.~\cite{Nakahara:1999vy,deForcrand:2000akx,Laermann:2001vg,Cheng:2010fe}.

\section{Discussion}
\label{sec:discussion}

The three dimensional CS matter systems exhibit various non-trivial
topological phases at low energy, and it has been conjectured that gauge
theories with fermions and bosons describe the same physics near the
critical point of the parameter spaces.
Although this duality is tightly related to the peculiar anyon
statistics in three space-time dimensions, the symmetry breaking
phenomena and dualities conjectured in QCD$_3$ 
with small CS levels
look quite similar to our QCD vacuum in four dimensions.

The winding $\theta$ background on an $S^1$ compactified space can
directly relate the three and four dimensional theories by comparing the
low energy limits.
We find that in QCD$_4$ the chiral phase transition should happen at a
critical radius, and there can be a description of the phase transition
as the Higgs mechanism of the $U(N_f)$ gauge theory where the gauge
bosons are the vector mesons.

Under the winding $\theta$, the WZ terms in the chiral Lagrangian leave
non-trivial WZ terms in the three dimensional effective theories, where
the baryon number can be identified as the magnetic flux of QED.
The origin of this unusual relation between baryons and monopoles can be
understood as the 't~Hooft anomaly in QCD$_4$.

We left the discussion of how to test the possibility of the vector
mesons becoming gauge bosons. One of the natural frameworks to discuss
this point is the holographic QCD where the vector mesons are already
gauge bosons in the Higgs phase. The specific holographic model such as
the Sakai-Sugimoto model~\cite{Sakai:2004cn} may be able to be used to
study the dynamics of the phase transition in the winding $\theta$
background. Also, if the feature of gauge bosons getting ``massless'' in
the four dimensional language remains in the trivial $\theta = 0$
background, the lattice QCD may be able to directly test the scenario.

Another non-trivial prediction of the model is the existence of the
monopoles which cut the string made of $\rho$ and
$\omega$~\cite{Kitano:2012zz}. 
We are not sure what should be identified as these objects in the hadron spectrum.
Due to the color-flavor locking, the
monopoles really carries the magnetic charge of QED while they are
confined by the string. It is certainly interesting to look for the
candidates of hadrons which are made of the monopole-string system.

\section*{Acknowledgements}
We would like to thank Ofer Aharony, Andreas Karch, Zohar Komargodski,
Yutaka Sakamura and Adi Armoni for discussions and useful comments.
RK and SY would like to thank the theory group at SLAC for hospitality
during their stay. RK also thanks the theory group at UC Davis for
warm hospitality during his stay.
The work of RK is supported by JSPS KAKENHI Grant No.~15KK0176,
19H00689, and MEXT KAKENHI Grant No.~18H05542.
The work of SY is supported in part by a center of excellence supported
by the Israel Science Foundation (2289118) and the Israel-Germany
Foundation (GIF). 

\appendix

\section{Lagrangian of quiver diagram}
\label{sec:quiver}

Here, we show the Lagrangian that is described by the quiver diagram in 
the left panel of Fig.~\ref{fig:quiver}:
\begin{equation}
\begin{split}
{\cal L}
=&i\bar{q}^{(L)0}\gamma^M(\partial_M-iA_M^{(L)})P_Lq^{(L)0}+i\bar{q}^{(R)0}\gamma^M(\partial_M-iA_M^{(R)})P_Rq^{(R)0}\\
&+\sum_{i=1}^{n_L}i\bar{q}^{(L)i}\gamma^M(\partial_M-ib_M^{(L)i})q^{(L)i}+\sum_{i=1}^{n_R}i\bar{q}^{(R)i}\gamma^M(\partial_M-ib_M^{(R)i})q^{(R)i}\\
&-\sum_{i=1}^{n_L}\frac{1}{2g_i^{(L)2}}\tr\left(f_{MN}^{(L)i}f^{(L)iMN}\right)-\sum_{i=1}^{n_R}\frac{1}{2g_i^{(R)2}}\tr\left(f_{MN}^{(R)i}f^{(R)iMN}\right)\\
&+\tr|\partial_MU_{01}^{(L)}-iA_M^{(L)}U_{01}^{(L)}+iU_{01}^{(L)}b_M^{(L)1}|^2+\sum^{{n_L}-1}_{i=1}\tr|\partial_MU_{i,i+1}^{(L)}-ib_M^{(L)i}U_{i,i+1}^{(L)}+iU_{i,i+1}^{(L)}b_M^{(L)i+1}|^2\\
&+\tr|\partial_MU_{10}^{(R)}-ib_M^{(R)}U_{10}^{(R)}+iU_{10}^{(R)}A_M^{(R)}|^2+\sum^{{n_R}-1}_{i=1}\tr|\partial_MU_{i+1,i}^{(R)}-ib_M^{(R)i+1}U_{i+1,i}^{(R)}+iU_{i+1,i}^{(R)}b_M^{(L)i}|^2\\
&+\tr|\partial_M\Phi-ib_M^{(L){n_L}}\Phi+i\Phi b_M^{(R){n_R}}|^2\\
&-\sum_{i=0}^{n_L-1}m_{i,i+1}^{(L)}\bar{q}^{(L)i+1}U_{i,i+1}^{(L)\dagger}P_Lq^{(L)i}-\sum_{i=0}^{n_L-1}m_{i+1,i}^{(R)}\bar{q}^{(R)i}U_{i+1,i}^{(R)\dagger}P_Lq^{(R)i+1}+{\rm h.c.}\\
&-m_\Phi\bar{q}^{(R){n_R}}\Phi^\dagger P_Lq^{(L){n_L}}+{\rm h.c.},
\end{split}
\end{equation}
where $n_L$ and $n_R$ are the number of double circle nodes in the upper and lower lines, respectively.

\section{Winding $\theta$ term}
\label{sec:windingTheta}

Here, we review how to treat the $\theta$ term with a winding
number~\cite{Cordova:2019jnf}.  When $\theta$ has winding number along a
compact direction, the $\theta$ term is not well defined on one
patch.

Let us consider an integral
\begin{align}
 {1 \over 2 \pi} \int_{S^1} \theta d q ,
\end{align}
with
\begin{align}
 \int_{S^1} d \theta = 2\pi k, \quad
 \int_{S^1} d q = 2\pi n, \quad (k,n \in {\mathbb Z}).
\end{align}
We would like to define the integral up to $2 \pi m$, $(m \in {\mathbb
Z})$ since the integral will be exponentiated in the path integral.
In Ref.~\cite{Cordova:2019jnf}, a general
prescription to define such an integral is discussed. By taking $t$, $(0
\le t < 2 \pi)$ as the coordinate on $S^1$, the prescription gives
\begin{align}
 &\fr{1}{2\pi} \int dt \theta(t) \dot{q} (t)
:=
\fr{1}{2\pi} 
\int_{0}^{2 \pi} 
dt \theta (t) \dot{q} (t)
 - k q(2 \pi).
\label{eq:intdef}
\end{align}
Similarly, the integration of $\dot{\theta}(t) q(t)$ can be defined by
\begin{align}
 &\fr{1}{2\pi} \int dt \dot\theta(t) q (t)
:=
\fr{1}{2\pi} 
\int_{0}^{2 \pi} 
dt \dot\theta (t) q (t)
 - n \theta (2 \pi).
\label{eq:intdef2}
\end{align}
The definitions of the integral in \ers{eq:intdef} 
and \eqref{eq:intdef2} have the following
desired features.
The integral is invariant modulo $2 \pi$ under the shifts, $\theta \to
\theta + 2\pi$ and $q \to q + 2 \pi$. Also, the integral does not depend
(modulo $2 \pi$) on the choice of the $t=0$ point on $S^1$.

These definitions are consistent with the integration by parts
(modulo $2\pi$), {\it i.e.},
\begin{align}
 \fr{1}{2\pi} \int dt \theta(t) \dot{q} (t)
&=
\fr{1}{2\pi} 
\int_{0}^{2 \pi} 
dt  \theta (t) \dot{q}(t) + k q (2\pi)
=
-\fr{1}{2\pi} 
\int_{0}^{2 \pi} 
dt \dot \theta (t) q(t) + n \theta (0)
\nonumber \\
&=
-\fr{1}{2\pi} 
\int 
dt \dot \theta (t) q(t), \mod 2 \pi.
\end{align}

\section{Baryon number and the configuration of ${\eta'}$}
\label{sec:baryon}

We discuss a configuration to give the baryon number, $B \neq 0$, in QCD
under a non-trivial background of ${\eta'}$. We now take the space-time as
the Minkowski space, $M^4$. The WZ term in QCD contains the following
term:
\begin{align}
 S_{\rm WZ} = - {N \over 8 \pi^2}
\int {\rm Tr} \left(
A dA + {2 \over 3} A^3 
\right) 
d {\eta'},
\label{eq:WZapp}
\end{align}
where the $N_f \times N_f$ matrix $A$ is the external gauge field which
couples to the $U(N_f)$ vector current, and ${\eta'}$ is the $U(1)$ part of
the Nambu-Goldstone mode, $U = e^{i \pi^a T^a + i {\eta'}}$.
The trace part of the gauge field normalized as, $B = (N / N_f) {\rm Tr}
A$, is the source for the baryon number. The baryon charge density can
be read off by differentiating with respect to $B$ as
\begin{align}
\rho_B = {1 \over 4 \pi^2} 
\epsilon_{ijk} {\rm Tr}(\partial_i A_j) \partial_k {\eta'} + \cdots.
\label{eq:baryon_density}
\end{align}

Let us consider a configuration with ${\eta'} = 0$ at $x_3 = -\infty$ and
${\eta'} = 2 \pi$ at $x_3 = + \infty$. We also apply an external magnetic
field of the (11) component of $A$ in the $x_3$ direction, with 
\begin{align}
 \int
dA^{(11)} = 2 \pi,
\end{align}
{\it i.e.,} putting a monopole and an anti-monopole
at $x_3 = \mp \infty$.

This configuration provides $B=1$ as one can see from the baryon density
in Eq.~\eqref{eq:baryon_density}. One can also understand this as the
Witten effect.
For the (11) component of $A$, there is an effective $\theta$ term from
Eq.~\eqref{eq:WZapp},
\begin{align}
 S_{\theta} = - {N \over 8 \pi^2} \int A^{(11)} d A^{(11)} d {\eta'} ,
\end{align}
which varies as a function of $x_3$.
Therefore, as we move a monopole from $x_3 = - \infty$ to $x_3 =
+ \infty$, the monopole obtains the electric charge, $N$, to couple to
$A^{(11)}$ by the Witten effect.
Since $A^{(11)} = B/N + \cdots$, the dyon carries the baryon number
$B=1$. Although the magnetic field is eliminated by this move, the
baryon number remains.

Now we consider the situation that the change of the value of ${\eta'}$
happens in a finite region on the $(x_1, x_2)$-plane and at a localized
location in the $x_3$ coordinate. This sheet-like configuration is
called the Hall droplet in Ref.~\cite{Komargodski:2018odf}.
If a monopole goes through the Hall droplet, it becomes a dyon with
$B=1$ by the Witten effect. This means that if we put magnets on the
both sides of the droplet, the magnetic lines cannot just go through the
droplet. In order to let the monopole line to go through, one needs to
throw in a baryon.
Conversely, starting from a configuration where the magnetic lines are
penetrating the droplet, when the magnets are turned off or taken away,
the system should relax to a state with a finite baryon number.
It has been discussed in Ref.~\cite{Komargodski:2018odf} that the
excitation of the edge mode of the droplet corresponds to the baryon
state with spin $N/2$.
This state is a good candidate of the remnant of the system.

The flavor quantum numbers of systems can be read off as in the same way
as the baryon number.
Let us take the cases with $N=3$ and $N_f = 2$ as in real QCD, where
$A^{(11)}$ couples to the current of the up quark. The configuration of
the unit magnetic line of $A^{(11)}$ going through the droplet now has
the quantum number of the operator $uuu$, {i.e.,} it has the electric
charge $Q=2$.
This is indeed the same as the baryon discussed in
Ref.~\cite{Komargodski:2018odf}.

Let's consider another example where only one of the diagonal components
of the Nambu-Goldstone field has the non-trivial configurations:
\begin{align}
 (\pi^a T^a + {\eta'}) \Big|_{x_3=-\infty} = \left(
\begin{array}{cccc}
 0 & & &\\
 & 0 & &\\
 & & \ddots &\\
 & & & 0\\
\end{array}\right), \quad
 (\pi^a T^a + {\eta'}) \Big|_{x_3=+\infty} = \left(
\begin{array}{cccc}
 2 \pi & & &\\
 & 0 & &\\
 & & \ddots &\\
 & & & 0\\
\end{array}\right).
\end{align}
When the $U(1)$ baryon (or QED as in the real world) is gauged, the
minimal magnetic charge is
\begin{align}
\int {d B \over N} = \int d A^{(11)} = \cdots = \int d A^{(N_f N_f)} = {2 \pi \over N},
\end{align}
as in the well-known magnetic monopole in grand unified
theories~\cite{tHooft:1974kcl, Polyakov:1974ek}. The Dirac quantization
conditions for quarks are satisfied by taking into account the ${\mathbb
Z}_N$ magnetic flux of $SU(N)$ carried by the
monopole~\cite{Englert:1976ng, Goddard:1976qe}.
The configuration that the magnetic line of this monopole, {\it i.e.,}
the 't~Hooft line, penetrates the Hall droplet has now the baryon number
$B=1/N$.
Therefore, once the magnets are removed, the system should relax to a
quark!
Again by the Witten effects, the 't~Hooft line accompanies a Wilson line
when it goes across the droplet. When we turn off the magnetic part, the
Wilson line which ends on the sheet remains. There should be a quark at
the end point of the Wilson line.
Indeed, there is an anyon excitation of the Hall droplet with the baryon
number $B=1/N$. (See \cite{Tong:2016kpv} for a review.) It is
interesting that the quark is described as a soliton made of hadrons!

The discussion here is closely related to the chiral soliton lattice
studied in Ref.~\cite{Brauner:2016pko}, 
where 
the pions develop winding numbers
under strong magnetic fields and a chemical potential of baryons.
Microscopically, one baryon can be converted into a configuration of a
Hall droplet with one unit of the magnetic flux penetrating through
it.

\bibliography{yokokura}
\end{document}